\documentclass{aa}

\usepackage{graphicx}
\usepackage{txfonts}

\usepackage{amsmath}
\usepackage{makecell}
\usepackage{csquotes}
\usepackage{booktabs}
\usepackage{natbib,twoopt}
\usepackage[breaklinks=true]{hyperref} 
\bibpunct{(}{)}{;}{a}{}{,}            



\usepackage{hyperref}
\hypersetup{
    colorlinks,
    linkcolor=blue,
    citecolor=blue, 
    urlcolor=blue
}
\begin{document}

   \title{Rapid, out of equilibrium metal enrichment indicated by a flat mass-metallicity relation at $z\sim 6$ from NIRCam grism spectroscopy}
   \titlerunning{MZR at $z\sim6$ using NIRCam grism}

   \author{Gauri~Kotiwale\inst{\ref{inst: ista}}\fnmsep\thanks{Email: gauri.kotiwale@ista.ac.at}
          \and
          Jorryt~Matthee\inst{\ref{inst: ista}}
          \and 
          Daichi~Kashino\inst{\ref{inst: NAOJ}}
          \and
          Aswin~P.~Vijayan \inst{\ref{inst: sussex}}
          \and
          Alberto~Torralba \inst{\ref{inst: ista}}
          \and
          Claudia~Di~Cesare\inst{\ref{inst: ista}}
          \and
          Edoardo~Iani\inst{\ref{inst: ista}}
          \and
          Rongmon~Bordoloi \inst{\ref{inst: north_carolina}}
          \and 
          Joel~Leja \inst{\ref{inst: psu1},\ref{inst: psu2}}
          \and 
          Michael~V.~Maseda \inst{\ref{inst: UoWisc}}
          \and
          Sandro~Tacchella \inst{\ref{inst: UoC_kavli},\ref{inst: UoC}}
          \and 
          Irene~Shivaei \inst{\ref{inst: CAB}}
          \and 
          Kasper~E.~Heintz \inst{\ref{inst: dawn},\ref{inst: NBI}, \ref{inst: UoGeneva}}
          \and 
          A.~Lola~Danhaive \inst{\ref{inst: UoC_kavli},\ref{inst: UoC}}
          \and
          Sara~Mascia\inst{\ref{inst: ista}}
          \and
          Ivan~Kramarenko\inst{\ref{inst: ista}}
          \and
          Benjam\'in~Navarrete\inst{\ref{inst: ista}}  
          \and
          Ruari~Mackenzie \inst{\ref{inst: EPFL}}
          \and
          Rohan~P.~Naidu \inst{\ref{inst: MIT}}
          \and
          David~Sobral \inst{\ref{inst: BNP}}}

   \institute{
        Institute of Science and Technology Austria, Am Campus 1, 3400 Klosterneuburg, Austria \label{inst: ista}
    \and 
        National Astronomical Observatory of Japan, 2-21-1 Osawa, Mitaka, Tokyo 181-8588, Japan \label{inst: NAOJ}
    \and 
        Astronomy Centre, University of Sussex, Falmer, Brighton BN1 9QH, UK \label{inst: sussex}
    \and 
        Department of Physics, North Carolina State University, Raleigh, 27695, North Carolina, USA \label{inst: north_carolina}
    \and
        Institute for Computational \& Data Sciences, The Pennsylvania State University, University Park, PA 16802, USA \label{inst: psu1}
    \and 
        Institute for Gravitation and the Cosmos, The Pennsylvania State University, University Park, PA 16802, USA \label{inst: psu2}
    \and 
        Kavli Institute for Cosmology, University of Cambridge, Madingley Road, Cambridge, CB3 0HA, UK \label{inst: UoC_kavli}
    \and 
        Cavendish Laboratory, University of Cambridge, JJ Thomson Avenue, Cambridge, CB3 0HE, UK \label{inst: UoC}
    \and
    Department of Astronomy, University of Wisconsin-Madison, 475 N. Charter St., Madison, WI 53706, USA \label{inst: UoWisc}
    \and 
    Centro de Astrobiolog\'ia (CAB), CSIC-INTA, Carretera de Ajalvir km 4, Torrej\'on de Ardoz, E-28850, Madrid, Spain \label{inst: CAB}
    \and
        Cosmic Dawn Center (DAWN), Denmark \label{inst: dawn}
    \and
        Niels Bohr Institute, University of Copenhagen, Jagtvej 128, 2200 Copenhagen N, Denmark \label{inst: NBI}
    \and
        Department of Astronomy, University of Geneva, Chemin Pegasi 51, 1290 Versoix, Switzerland \label{inst: UoGeneva}
    \and
        Laboratory of Astrophysics, \'Ecole Polytechnique F\'ed\'erale de Lausanne (EPFL), Observatoire de Sauverny, 1290 Versoix, Switzerland \label{inst: EPFL}
    \and
        MIT Kavli Institute for Astrophysics and Space Research, Massachusetts Institute of Technology, Cambridge, MA 02139, USA \label{inst: MIT}
    \and
        BNP Paribas Corporate \& Institutional Banking, Torre Ocidente Rua Galileu Galilei, 1500-392 Lisbon, Portugal \label{inst: BNP}
    }

   \date{Received August 2025}

\abstract{
We aim to characterise the mass-metallicity relation (MZR) and the 3D correlation between stellar mass, metallicity and star-formation rate (SFR) known as the fundamental metallicity relation (FMR) for galaxies at $5<z<7$. Using $\sim800$ [O\,{\sc iii}] selected galaxies from deep NIRCam grism surveys, we present our stacked measurements of direct-$T\rm_e$ metallicities, which we use to test recent strong-line metallicity calibrations. Our measured direct-$T\rm_e$ metallicities ($0.1$--$0.2\,\rm Z_\odot$ for M$_\star$ $\approx5\times10^{7-9}$ M$_{\odot}$, respectively) match recent JWST/NIRSpec-based results. However, there are significant inconsistencies between observations and hydrodynamical simulations. We observe a flatter MZR slope than the SPHINX$^{20}$ and FLARES simulations, which cannot be attributed to selection effects. With simple models, we show that the effect of an [\ion{O}{iii}] flux-limited sample on the observed shape of the MZR is strongly dependent on the FMR. If the FMR is similar to the one in the local Universe, the intrinsic high-redshift MZR should be even flatter than observed. In turn, a 3D relation where SFR correlates positively with metallicity at fixed mass would imply an intrinsically steeper MZR. Our measurements indicate that metallicity variations at fixed mass show little dependence on the SFR, suggesting a flat intrinsic MZR.
This could indicate that the low-mass galaxies at these redshifts are out of equilibrium and that metal enrichment occurs rapidly in low-mass galaxies. However, being limited by our stacking analysis, we are yet to probe the scatter in the MZR and its dependence on SFR. 
Large carefully selected samples of galaxies with robust metallicity measurements can put tight constraints on the high-redshift FMR and, help to understand the interplay between gas flows, star formation and feedback in early galaxies.
}

   \maketitle

\section{Introduction}
\label{sec: intro}

The gas-phase metallicity of the interstellar medium of galaxies reflects the interplay of gas inflows, feedback, and star formation \citep{2008MNRAS.385.2181F, 2013ApJ...772..119L, 2019A&ARv..27....3M, 2025arXiv250408933C}. Usually, the gas-phase metallicity is expressed in terms of the oxygen abundance (relative to the hydrogen abundance) as it is the most abundant metal in the Universe. It is most easily observed through strong lines in star-forming regions of galaxies, such as [\ion{O}{ii}]$\lambda\lambda 3726,3729$ and [\ion{O}{iii}]$\lambda\lambda4960,5008$. Studying the evolution of the gas-phase metallicities of galaxies through cosmic time can tell us about the baryonic mass assembly in the Universe.

\par
The relation between gas-phase metallicity ($Z_{\rm gas}$) and stellar mass (M$_\star$), known as the mass-metallicity relation (MZR), has been extensively studied, spanning from galaxies in the local Universe \citep[at $z\sim0$, e.g.][]{2004ApJ...613..898T, 2008ApJ...672L.107E, 2011arXiv1112.3300M, AandM_FMR2013, Curti_FMR2020, Scholte2024_lowz}, to those at the peak of cosmic star formation activity (up to $z\sim 3$) \citep[e.g.][]{2005ApJ...635.1006S,2008A&A...488..463M,2009MNRAS.398.1915M,2014ApJ...789L..40W,2021MNRAS.506.1237T,Sanders2021MZR}, and well into to epoch of reionisation using JWST \citep[e.g. ][]{ 2023NatAs...7.1517H,2023ApJ...950...67M, 2023ApJS..269...33N, Chemerynska2024, 2024A&A...681A..70L, Curti2024MZR, Sanders2024_MZR, Sarkar2025_MZR, Chakraborty2025_MZR, 2024ApJ...971...43M, 2025MNRAS.540.2176C, 2025arXiv250418616L, Pollock2025_MZR}. It has been observed that the gas-phase metallicity of galaxies increases with increasing stellar mass, and for $z<4$ the slope of this trend becomes shallower at higher masses. This trend is explained through gas regulation in galaxies \citep{ 2008MNRAS.385.2181F,PeeplesShankar2011,2012MNRAS.421...98D,2013ApJ...772..119L}. Due to higher stellar masses, the galaxies have a deeper potential well, making the outflow of metal-enriched gas from the galaxy difficult. The metallicity also decreases at fixed stellar mass with increasing redshift. The redshift evolution of the MZR could be due to inflow of metal-poor gas, or outflow of metal-rich gas due to stronger feedback mechanisms at higher redshifts \citep{Lopez_FMR2010, 2013MNRAS.433.1425B, 2016A&A...595A..48B, Scholte2024_lowz}. This evolution has also been studied in terms of the 3D relation between gas-phase metallicity, stellar mass and star-formation rate \citep{Mannucci_FMR2010,Lopez_FMR2010,AandM_FMR2013,Curti_FMR2020}. The scatter in the MZR in the local Universe can be reduced by accounting for variations in the star formation rates (SFR) of galaxies. Galaxies in the local Universe are found to reside on a 2D plane in the M$_\star$-$Z_{\rm gas}$-SFR space with $<0.1$ dex scatter. This relation is known as the fundamental metallicity relation (FMR). Multiple studies find that the FMR remains invariant up to $z\sim3$ \citep{2012MNRAS.421..262C, 2015A&A...577A..14M, 2015PASJ...67..102Y, 2021MNRAS.506.1237T, Sanders2021MZR}. \cite{2023ApJS..269...33N} and \cite{Sarkar2025_MZR} find no redshift evolution for their sample between $4<z<8$ compared to the $z=0$ FMR, but observe significant deviations at $z>8$. Various theoretical models and simulations also explore the FMR in terms of SFR on different spatial scales \citep{2021ApJ...910..137W} and temporal scales \citep{Torrey2018_FMR, 2024ApJ...971L..14M}, the impact of different feedback mechanisms \citep{2017MNRAS.472.3354D}, and variations in gas flows and gas fraction \citep{2013MNRAS.430.2891D, 2019MNRAS.484.5587T}.\par
There are several arguments about whether the FMR exists out to high-redshifts \citep[e.g.][]{2014ApJ...792...75Z, 2023NatAs...7.1517H, Curti2024MZR, Scholte2025_EXCELS, 2025MNRAS.540.2176C}. If truly fundamental, the shape of the FMR plane does not evolve -- only the relative positions of the galaxy population on the plane shifts. Simulations suggest that the existence of the plane depends on the time-scales of SFR variability \citep[e.g.][]{Torrey2018_FMR, 2024ApJ...971L..14M}. Given that the star formation histories of high-$z$ galaxies appear more bursty \citep[e.g.][]{Topping2022_burstySFH, Endsley2023_burstySFH,Looser2025_burstySFH}, it could cause the FMR to break down at high-$z$. Indeed, some JWST studies indicate that the normalisation of the plane shifts, i.e. galaxies systematically have lower metallicities than predicted by the $z=0$ FMR \citep{2023NatAs...7.1517H, Curti2024MZR, Scholte2025_EXCELS}.

\par
Gas-phase metallicity is typically inferred using two approaches: (1) the direct-$\rm T_e$ method, (2) strong-line calibrations.  The direct-$\rm T_e$ method is based on electron temperature ($\rm T_e$) determination using auroral emission lines. The most common auroral line used to determine the electron temperature of the $O^{+2}$ regions is [\ion{O}{iii}]$\lambda 4364$ \citep{2019A&ARv..27....3M, 2025arXiv250408933C}. The relative strength of the [\ion{O}{iii}]$\lambda 4364$ and [\ion{O}{iii}]$\lambda 5008$  lines can be used to calculate the electron temperature. The calculated electron temperature is then used to estimate the gas-phase metallicity using dust-corrected emission line ratios of [\ion{O}{iii}] and [\ion{O}{ii}] with H$\beta$. The challenge with this method is that the [\ion{O}{iii}]$\lambda 4364$ auroral line is very faint and thus hard to detect in most galaxies \citep{2006agna.book.....O, 2019ARA&A..57..511K}.  
\par
On the other hand, the strong line method is applicable to a larger dataset as it only requires the detection of emission lines that are easily observable in the star-forming regions of the galaxies.
The strong line metallicity calibrations are usually calibrated to metallicity measurements from the direct-$\rm T_e$ method. Multiple studies have constructed metallicity calibration curves using the observed relationship between emission line ratios and the direct-Te method metallicity estimates for galaxies \citep[e.g.][]{2008A&A...488..463M, 2013A&A...559A.114M, AandM_FMR2013,2018ApJ...859..175B, Curti_FMR2020,Nakajima2022,2024A&A...681A..70L,Chakraborty2025_MZR, Sanders2024_MZR,Scholte2025_EXCELS}. Some of the calibration curves are developed using galaxies which are considered as high-redshift analogues \citep{2018ApJ...859..175B,Nakajima2022}. These calibrations offer a potential solution to estimate metallicity at high-$z$ where getting direct-T$\rm _e$ metallicities is difficult, however their applicability at these redshifts needs to be tested.

\par
Recent studies using JWST make use of strong-line calibrations to explore the MZR at a redshift range of $z=3-10$ \citep{2023ApJ...950...67M, 2023ApJS..269...33N, Li23_z2_3,2023NatAs...7.1517H, Curti2024MZR, Sarkar2025_MZR,  Chemerynska2024, Rowland2025_Rebels, 2025arXiv250418616L,Cataldi2025_marta,Pollock2025_MZR}. Studies by \cite{2023ApJS..269...33N, Sanders2024_MZR, 2024A&A...681A..70L, Sarkar2025_MZR,Chakraborty2025_MZR},  and \cite{Pollock2025_MZR} use the [\ion{O}{iii}]$\lambda4364$ auroral line (direct-$\rm T_e$ method) to investigate the MZR up to $z\sim 10$. Most of these studies mainly focus on galaxies with log(M$_\star /$M$_{\odot})>8$. Analysis by \cite{Curti2024MZR} and \cite{Chemerynska2024} explore the MZR in down to $ \rm log(M_\star/M_{\odot})\sim 6$   at high redshifts. Considering JWST's recent arrival on the scene, the sample sizes for which we have direct-$\rm T_e$ measurements are still limited. 

\par
In this work, we study and characterise the MZR and the 3D correlation with SFR
between $5 < z <7$  through Direct-T$_e$ metallicities using data from deep NIRCam grism surveys EIGER \citep{2023ApJ...950...66K}, COLA1 \citep{2024A&A...689A..44T} and ALT \citep{2024arXiv241001874N}. Our sample makes use of NIRCam spectra, compared to most literature studies that use JWST/NIRSpec. Grism spectra have a simple flux-limited selection function and are not subject to slit losses. On the other hand, it is less sensitive and  has a smaller wavelength coverage than NIRSpec spectra. Our galaxy sample is selected by using only the [\ion{O}{iii}]$\lambda \lambda 4960,5008$ flux in all three surveys, giving us a well defined, flux-limited selection function.  We cover the H$\beta$ and [\ion{O}{iii}]$\lambda \lambda 4960,5008$ emission lines for all our galaxies and, for a subset, we also cover H$\gamma$ and [\ion{O}{iii}]$\lambda4364$. For this subset we measure the [\ion{O}{iii}]$\lambda4364$ auroral line with SNR $>3$ in stacks of stellar mass bins, which we use to calculate the direct-T$\rm_e$ metallicities. We use these measurements to test the strong-line calibrations high-$z$ galaxies, obtain a scaling relation between the gas-phase metallicity and stellar mass, and investigate the impact of our selection function.

\par
The paper is structured as follows: we describe our NIRCam grism dataset and data reduction pipeline in Sect. \ref{sec: data}. We report the method for extracting emission line fluxes and dust correction in Sect. \ref{sec: methods}. In Sect. \ref{sec: measurements}, we describe our methods to derive metallicities using both, the direct-Te method and strong-line calibrations and, in Sect. \ref{sec: MZR}, we present our MZR and make comparisons to the literature. Furthermore, we discuss the impact of our selection function in Sect. \ref{sec: discussion} through comparisons to simulations and toy models. Finally, we present our conclusions and summary in Sect. \ref{sec: summary}. \par 
Throughout this work, we adopt a flat $\Lambda$CDM cosmology with $\Omega_{\Lambda} = 0.69$,  $\Omega_{M} = 0.31$, and $H_0$ = 67.7 km s$^{-1}$ Mpc$^{-1}$ \citep{2020A&A...641A...6P}. All magnitudes are mentioned in the AB system \citep{AB_mags_1983}.

\section{Data}
\label{sec: data}
\subsection{Observations}
\label{subsec: observations}
We use a compilation of imaging and Wide Field Slitless Spectroscopy (WFSS) from deep NIRCam grism surveys. The data are obtained from the EIGER survey \citep[program ID 1243, PI Lilly,][]{2023ApJ...950...66K}, the ALT survey \citep[program ID 3516, PIs Matthee and Naidu,][]{2024arXiv241001874N} and the JWST program ID 1933 (PIs Matthee and Naidu, hereafter `COLA1 survey'). The EIGER survey is a deep grism cycle 1 survey in the F356W filter centred on six widely separated $z>6$ quasar fields. It contains imaging in the short-wavelength (SW) filters F115W and F200W, and direct imaging in the F356W long-wavelength (LW) filter. Each field has a survey area of 6.5$\times$3.4 arcmin$^2$ (i.e. 132 arcmin$^2$ in total). 

The WFSS in the F356W LW filter covers an observed wavelength of $3.1-4.0\mu$m with a spectral resolution of $R \sim 1600$ for a point source. The COLA1 survey has a similar observing strategy to EIGER, but is centred on the bright double-peaked Ly$\alpha$ emitter, COLA1 at $z=6.6$ \citep{2018A&A...619A.136M,2024A&A...689A..44T}. The total area in the COLA1 field is 21 arcmin$^2$. Finally, the ALT survey is the deepest NIRCam WFSS survey to date, targeting a wider 30 arcmin$^2$ area around the gravitational lensing cluster Abell 2744 in the F356W filter. Thanks to deep imaging from UNCOVER \citep[program ID 2561, PI Labbe,][]{2024ApJ...974...92B} and MegaScience \citep[program ID 4111, PI Suess,][]{2024ApJ...976..101S} there is excellent photometric coverage allowing detailed characterisation of the SEDs for galaxies from the ALT survey. From ALT, we only select galaxies with magnification factor $(\mu)<3$ to remove multi-imaged sources and strong arcs.
\par

\subsection{Data reduction}
\label{subsec: data_red}

In this work, we focus on [\ion{O}{iii}]$\lambda \lambda 4960, 5008$ selected galaxies which are observed in the NIRCam WFSS with the F356W filter ($\lambda \sim 3.1-4.0 \mu$m) in the redshift range $5.5<z<7$. The analysis in our paper makes use of both JWST/NIRCam imaging and WFSS. The details of the basic reduction steps are described in \cite{2023ApJ...950...66K,Kashino25} for the EIGER survey, \cite{2024A&A...689A..44T} for the COLA1 survey, and \cite{2024arXiv241001874N} for the ALT survey. Here we briefly summarise the WFSS data reduction steps for these surveys and additional modifications made for emission-line spectra extraction. \par

The WFSS data have been reduced using a combination of the JWST pipeline and customised python-based processing. We download the {\sc rate} files from the MAST database\footnote{\url{https://mast.stsci.edu/}} and use the JWST pipeline to flat field images. For the images in the COLA1 and EIGER fields, we subtract the background by calculating the median value along each column (which is orthogonal to the dispersion direction). For the ALT data in the Abell 2744 cluster, the median statistic is boosted by diffuse light from the vast number of sources (in particular cluster galaxies). We therefore construct (for each module) a master-background template based on the backgrounds in the COLA1 and EIGER data and scale the mode of each ALT grism image to the mode value of the background template. We align the astrometry of each image by matching the WCS of the image taken simultaneously in the short-wavelength filter to the reference catalogue based on the direct imaging in the F356W filter. In our basic reduction used to identify the sources, we separated the dispersed light in a `continuum' and an `emission-line' image, by estimating the continuum in each row with a running median filter along the dispersion direction. The filter is a kernel with width 51 pixels and a central hole of 9 pixels, to avoid over-subtraction of emission-lines. As detailed in \cite{2023ApJ...950...66K}, this procedure was run twice,  masking identified emission-lines from the first iteration.

In this study, we present an improved method for the extraction of emission-line (EMLINE) spectra  to avoid over-subtraction around emission lines. This was performed already in \cite{Matthee24}, but here we generalise it for [\ion{O}{iii}] emitters. The standard methodology can lead to slight over-subtraction of faint wings that are not identified in individual images. Building upon the aforementioned works for the EIGER, COLA1 and ALT surveys, we know the location of the emission lines in our spectra which allows us to subtract the continuum optimally without over-subtracting line emission. We use this information and stack the EMLINE spectra of galaxies with a [\ion{O}{iii}]$\lambda$5008 signal-to-noise ratio (SNR) $> 25$. The methodology for stacking is described further in Sect. \ref{subsec: stacking}. Once we have the stacked EMLINE spectra we generate 2D Gaussian masks, with a FWHM of 7 pixels and size of 15 pixels with a threshold of $1\times 10^{-20} \rm erg\ s^{-1}\ cm^2 $, for the [\ion{O}{iii}]$\lambda \lambda 4960,5008$ emission lines using the \texttt{Photutils} Python package \citep{larry_bradley_2024_13989456}. The mask for the [\ion{O}{iii}]$\lambda$5008 line is applied to its respective location, while the mask for the [\ion{O}{iii}]$\lambda$4960 line is applied to the locations of [\ion{O}{iii}]$\lambda$4960, H$\gamma$, [\ion{O}{iii}]$\lambda4364$, and H$\beta$ in each 2D science (SCI) spectrum in our sample. We apply the mask for [\ion{O}{iii}]$\lambda4960$ to the other emission lines (H$\gamma$, [\ion{O}{iii}]$\lambda4364$, and H$\beta$) due to their relatively smaller flux compared to [\ion{O}{iii}]$\lambda$5008. Utilizing the [\ion{O}{iii}]$\lambda5008$ mask for these lines could result in over-masking which can cause under-subtraction of the continuum and can cause the potential overlap with adjacent emission lines, especially for the H$\gamma$ and [\ion{O}{iii}]$\lambda4364$ emission lines. After applying the masks to the SCI spectra, we perform the continuum subtraction on the SCI spectra using using a running box-car median with a kernel of 71 pixels and a gap of 7 pixels in the centre, similar to the method in the initial extraction of emission-line spectra. The parameters were iteratively chosen based on stacked spectra, where we optimised to match the line-shapes and the 1:2.98 normalisation of the [\ion{O}{iii}]$\lambda\lambda$4960,5008 doublet.  \par

\begin{figure}
   \centering
    \includegraphics[width=8cm]{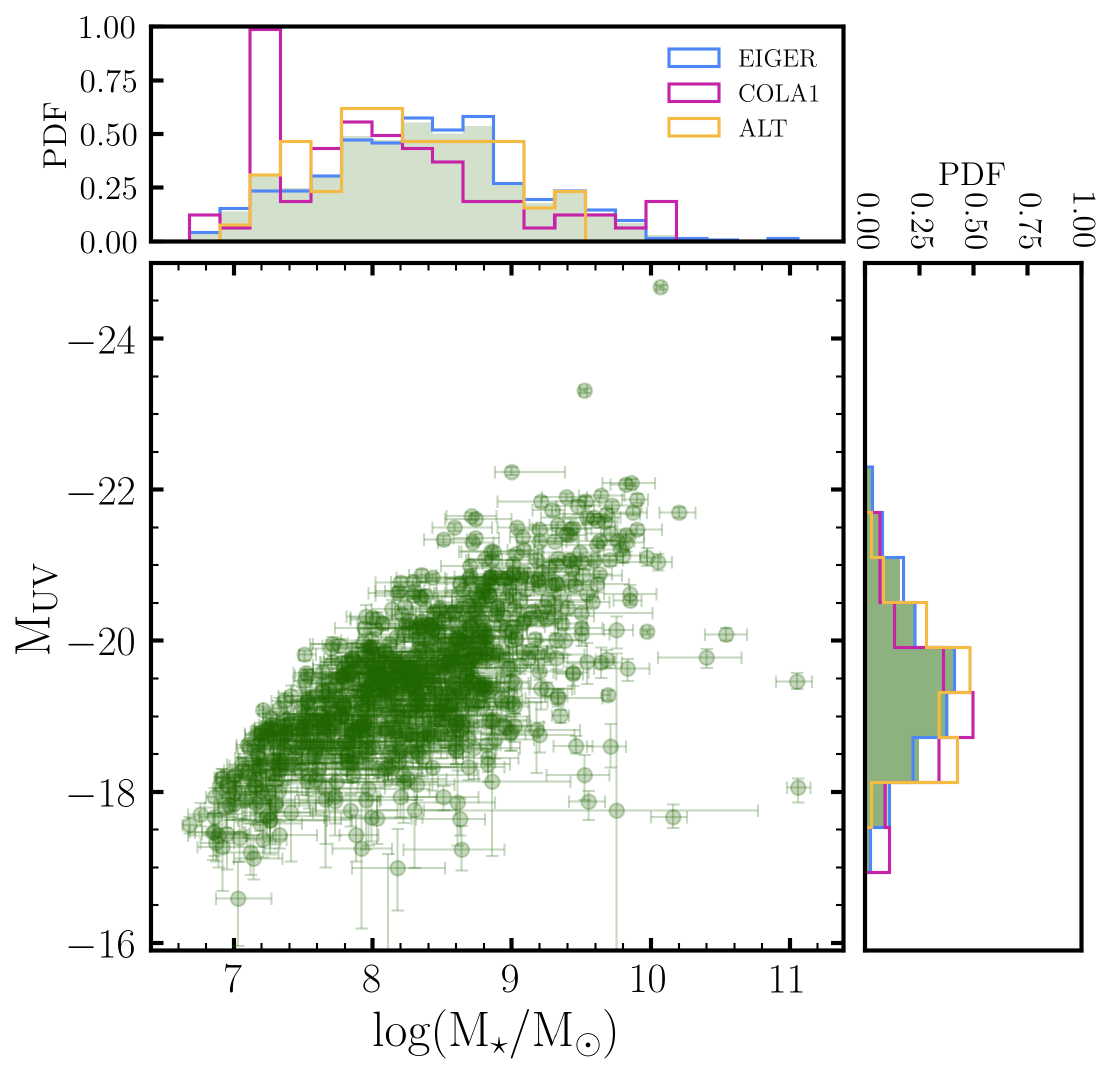}
      \caption{Relation between UV magnitude and stellar mass for our H$\beta$ sample (792 galaxies). The $\rm M_{UV}$ and stellar masses have been obtained through SED fitting using \texttt{Prospector}. With data from the ABELL2744 lensing field, we probe galaxies with $\rm M_{UV}$ down to $-16.6$ mag. The UV luminosity increases with stellar mass as expected.}
         \label{fig: MUVs}
   \end{figure}

\section{Methods}
\label{sec: methods}

\subsection{Galaxy sample}
\label{subsec: gal_sample}

Our dataset comprises of [\ion{O}{iii}]$\lambda \lambda 4960, 5008$ selected galaxies from the EIGER \citep{Kashino25}, COLA1 \citep{2024A&A...689A..44T} and ALT \citep{2024arXiv241001874N} surveys. These galaxies are selected using the `forward' method described in \cite{2023ApJ...950...66K}. In this method, the [\ion{O}{iii}] doublet sample are required to have $\rm S/N>3$ and a maximum offset of 3 pixels ($0.09''$) in their spatial centroid. Once we have our emission line spectra, the galaxies are inspected to detect any broad line H$\beta$ emission. We do this to find possible broad-line AGN candidates and remove them from our sample. We exclude the spectra for the central quasars in the EIGER fields. We also remove AGNs that are known from previous studies, such as the X-ray detected AGN, J1148$+$5253 \citep{2005ApJ...634L...9M} at $z = 5.69$. We also remove the galaxy J0100$-$9148 from our sample because of severe contamination in the 2D spectra. We do not find additional broad H$\beta$ emitters in our sample, however it should be noted that broad-line H$\alpha$ AGNs at high-redshifts usually have extremely faint broad H$\beta$ emission \citep{Brooks2025_dustyAGN}. Through this, we obtain a dataset of 792 galaxies, out of which there are 659 galaxies from the EIGER survey, 74 from the COLA1 survey and 59 from the ALT survey. We classify our data into two samples,
\begin{itemize}
    \item[(i)] H$\beta$ sample: galaxies for which our data cover the H$\beta$ emission line at rest-frame wavelength of 4862.68~\AA, i.e. galaxies with redshift $5.5\leq z<7$. This is the entire dataset with 792 galaxies.
    \\
    \item[(ii)] H$\gamma$ sample: galaxies for which our data cover the H$\gamma$ emission line at rest-frame wavelength of 4340~\AA, i.e. galaxies with redshift $6.3\leq z<7$. This is subset of the H$\beta$ sample, with 270 galaxies.

\end{itemize}

The typical UV magnitudes of the H$\gamma$ and the H$\beta$ sample are $M_{\rm UV}=$ $-19.5$ mag and $-19.4$ mag, respectively (see Fig. \ref{fig: MUVs}). Stellar masses in of our total dataset range from $\rm log(M_\star/M_\odot)= 6.9-11.1$, with typical stellar masses of both samples being $\sim 10^{8.3}$ $M_\odot$. Our sample is stellar mass complete down to $\rm log(M_\star/M_\odot) \sim 8.5$ (see Fig. \ref{fig: MUVs}). Properties such as stellar mass, emission line fluxes and photometry for the galaxies from the ALT catalogue have been corrected for their magnification using magnification factors from \cite{Furtak_magnifs_2023} and \cite{Price_magnifs_2025}. The redshift uncertainty for our sample is $\sim 60-70 \rm km\ s^{-1}$ \citep{2024A&A...689A..44T, 2024arXiv241001874N, Kashino25}.

\subsection{SED modelling}
\label{subsec: sed_model}

In order to estimate galaxy properties such as the stellar mass, we perform SED fitting using the code \texttt{Prospector} \citep{2021ApJS..254...22J}. Nebular emission is self-consistently modelled based on the photoionisation code \texttt{Cloudy} \citep{1998PASP..110..761F, 2013RMxAA..49..137F}. The stellar population modelling is implemented through \texttt{FSPS} \citep{2009ApJ...699..486C,2010ApJ...712..833C} using the MIST stellar models \citep{2016ApJS..222....8D, 2016ApJ...823..102C}. We use a non-parametric star-formation history with seven bins, adopting a `bursty continuity' prior. The first two bins are fixed at look-back times of 0-5 Myr and 5-10 Myr with the rest logarithmically spaced out to $z=20$. We adopt a \cite{2003PASP..115..763C} initial mass function (IMF), with a cut-off at 150 M$_\odot$. The other free parameters are the total stellar mass formed, stellar metallicity, gas-phase metallicity, dust attenuation (a screen and an additional component for young stars), and ionization parameter. The set-up for the SED fitting is described in detail in \cite{2022ApJ...940L..14N, 2024arXiv241001874N}, which is a modification of the prescription by \cite{2022ApJ...927..170T}. It should be noted that the derived stellar masses are strongly dependent on the assumed star formation history (SFH) model. \cite{Whitler2023_SFH} find that non-parametric models with continuity prior can infer stellar masses up to an order of magnitude larger (an order of magnitude smaller sSFR) than the constant SFH models at the ages of $ \lesssim 10$ Myr.\par
The SED modelling is done using photometric data and emission-line fluxes from the spectra. The photometric data used for SED fitting of galaxies from the EIGER, COLA1 and ALT survey varies due to differences in photometric filter coverage. For galaxies from the EIGER survey, we use photometry in the F115W, F200W and F356W filters. In the COLA1 survey, we have photometry in an additional band, F150W, which we include in the SED modelling. For galaxies from the ALT survey, we have photometry from numerous programs that targeted the Abell 2744 field (primarily UNCOVER; \cite{2024ApJ...974...92B} and MegaScience; \citealt{2024ApJ...976..101S}) totalling 27 filters available for the SED fitting\footnote{Wavelengths around the Lyman-$\alpha$ line are ignored in the SED fitting because of uncertainty in the IGM attenuation.}. For the galaxies from the EIGER and COLA1 surveys, we additionally use the H$\beta$ and the [\ion{O}{iii}]$\lambda\lambda4960,5008$ emission line fluxes derived from the grism spectra. For the ALT galaxies, the emission line fluxes are de facto included due to the photometry in the various medium-band filters. We have verified that there are no systematic biases between the stellar masses from the EIGER/COLA1 versus the ALT surveys by confirming that the relation between observed (magnification-corrected) F200W magnitude with the stellar mass is similar across the fields. We have also confirmed that in the ALT data the emission-line fluxes from grism data are consistent with those inferred from medium-band photometry (Mascia et al. in prep).

\subsection{Stacking Grism Spectra}
\label{subsec: stacking}
\begin{figure*}
   \centering
    \includegraphics[width=18cm]{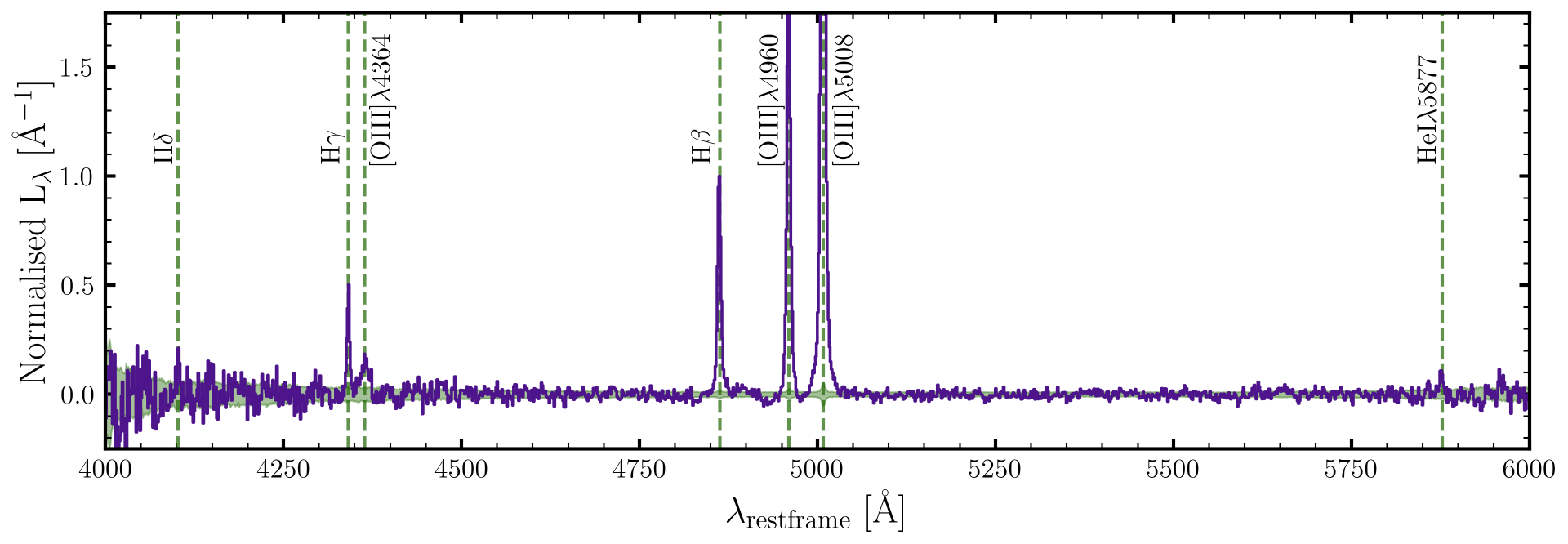}
      \caption{Median stacked 1D rest-frame emission-line spectrum of the full sample of 792 [\ion{O}{iii}] emitters at $z = 5.5 - 7$. Each galaxy spectrum has been normalised by its ([\ion{O}{iii}]$\lambda$5008$+2.98$[\ion{O}{iii}]$\lambda$4960)/2 flux before stacking. The green shaded region shows the uncertainty estimated through bootstrap resampling. We highlight the wavelengths of H$\delta$, H$\gamma$, [\ion{O}{iii}]$\lambda$4364, H$\beta$, [\ion{O}{iii}]$\lambda\lambda$4960,5008 and HeI$\lambda$5877.  For visualisation, we normalise the stacked spectrum by the H$\beta$ luminosity such that H$\beta$ peaks at unity.}
         \label{fig: stacked_spec}
   \end{figure*}

Emission lines in the spectrum of a galaxy are a probe of the physical conditions of the ISM. However obtaining a high signal-to-noise ratio for faint emission lines such as H$\gamma$ or [\ion{O}{iii}]$\lambda4364$ in individual galaxies is difficult with NIRCam grism data. In such cases, stacking methods can extract average measurements by reducing random noise fluctuations at the expense of losing information on individual sources and therefore variation across the stacked sample. For illustrative purposes, we show the 1D stacked spectra of all the galaxies in our dataset in Fig. \ref{fig: stacked_spec}. The 2D spectra have been brought to rest-frame and normalised by their ([\ion{O}{iii}]$\lambda$5008$+2.98$[\ion{O}{iii}]$\lambda$4960)/2 luminosity before stacking. red From the 2D stacked spectrum we extract the 1D spectrum as described in Sect. \ref{subsubsec: extracting_1D_spec}. We can see strong [\ion{O}{iii}]$\lambda\lambda 4960, 5008$ and H$\beta$ emission lines along with the H$\gamma$, H$\delta$ and [\ion{O}{iii}]$\lambda4364$. We also see the HeI-5877 emission line in the stacked spectrum, albeit relatively faint. In comparison to the stacks presented in \cite{2024ApJ...976..193R} for galaxies between $z=5-7$, we detect almost all the key emission lines over the same wavelength range (4000 to 6000~\AA). Additionally, we resolve the [\ion{O}{iii}]$\lambda \lambda 4960, 5008$ doublet, and H$\gamma$ and [\ion{O}{iii}]$\lambda4364$ due to the higher resolution of NIRCam/WFSS spectra in the F356W filter ($R\sim$1600). Although our grism spectra has higher resolution, our wavelength coverage is limited as compared to NIRSpec/Prism spectrum, which covers from 1000 to 7000 \AA~ in the same redshift range.\par
\begin{figure*}
   \centering
   \begin{tabular}{cc}
    \includegraphics[width=12cm]{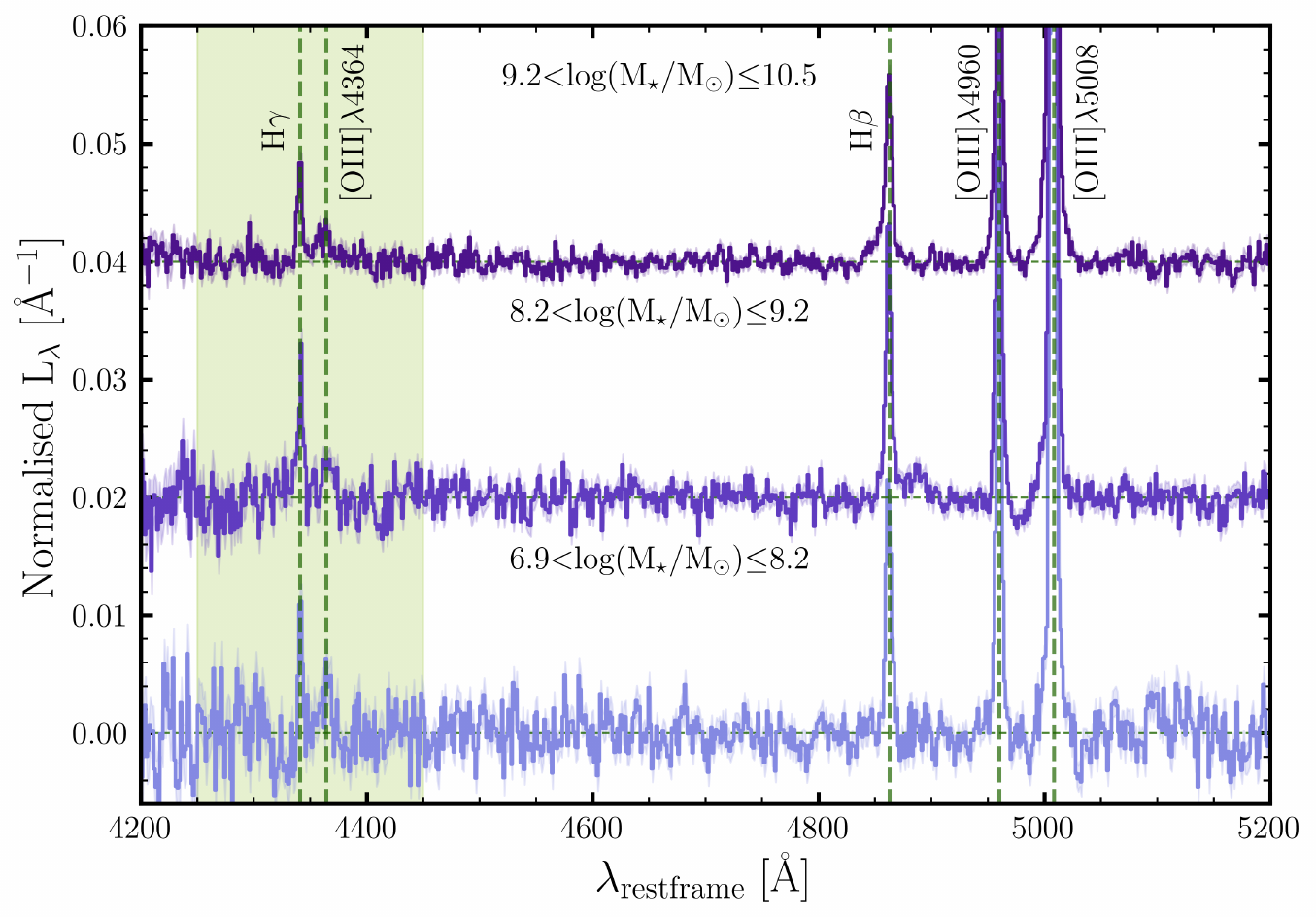} & 
        \includegraphics[width=4cm]{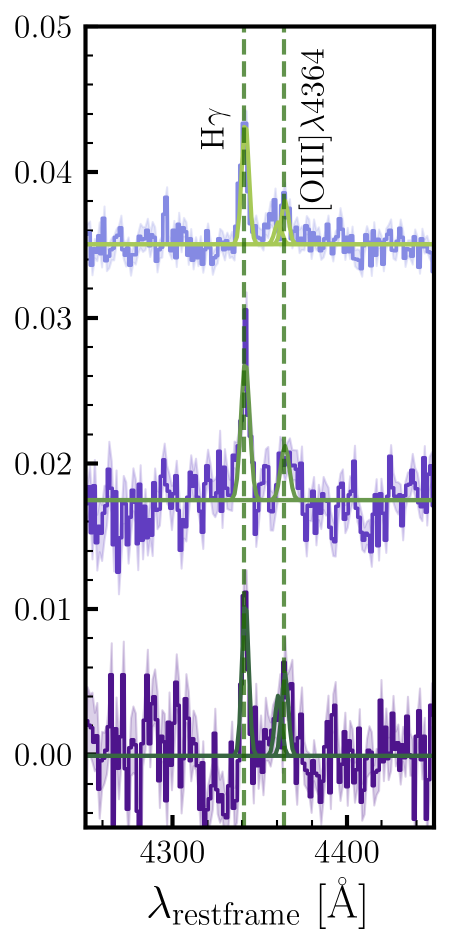} \\

    \end{tabular}
      \caption{Median stacked 1D rest-frame emission-line spectra of the H$\gamma$ sample ($z = 6.3 - 7$), in three stellar mass bins offset for visualisation purposes. The shaded region shows the uncertainty estimated through bootstrap resampling. In the right panel, we zoom into the wavelength range containing  H$\gamma$ and [\ion{O}{iii}]$\lambda$4364. We fit a complex of three Gaussians with the same standard deviation to obtain the fluxes of  H$\gamma$ and [\ion{O}{iii}]$\lambda$4364. An additional Gaussian component is fitted to account for possible contamination by [\ion{Fe}{ii}]$\lambda4360$. }
         \label{fig: Hgamma_stacked_spec}
   \end{figure*}

After dividing our data into $\rm H\gamma$ ($6.3\leq z<6.9$) and $\rm H\beta$ ($5.5\leq z<6.9$) samples and obtaining our emission-line spectra, as described in Sect. \ref{sec: data}, we perform median stacking of the emission-line spectra in stellar mass bins. In order to stack the spectra, we shift the 2D spectrum of each galaxy to the rest-frame wavelength grid ranging from 4000 to 6000 ~\AA~ through linear interpolation resampling. We have verified that our resampling is flux-conserving. We then normalise the 2D spectrum of each galaxy by its respective ([\ion{O}{iii}]$\lambda5008$ $+ 2.98 \cdot $[\ion{O}{iii}]$\lambda4960$)/2 luminosity. This allows us to ensure that the stacks represent the typical line ratios of the spectra and are not dominated by more luminous galaxies. We obtain the stacked spectra by calculating the median over the 300 bootstrap (with replacement) realisations. The uncertainties are obtained by calculating the standard deviation of these bootstraps. The median-stacked spectra of our H$\gamma$ sample is shown in Fig. \ref{fig: Hgamma_stacked_spec}.

\begin{table*}
\centering
\caption{Metallicity and SFR measurements in bins of stellar mass for the H$\gamma$ sample with $6.3<z<7$. }

\begin{tabular}{ccccc}
\midrule
\multicolumn{1}{c}{Mass Bin}                                  & No.~ of Galaxies & log$\rm (M_{\star}/M_{\odot})$ \tablefootmark{a}& 12+log(O/H) \tablefootmark{b}& log$_{10}$(SFR / $\rm M_\odot\ yr^{-1}$) \tablefootmark{c}\\ \midrule
\multicolumn{1}{c}{$6.89\leq$log(M$_\star$/M$_\odot$)$<8.2$}  & 113  & $7.77^{+0.32}_{-0.46}$& $7.69^{+0.17}_{-0.13}$&  $0.48\pm0.12$\\[2pt]
\multicolumn{1}{c}{$8.2<$log(M$_\star$/M$_\odot$)$<9.2$}      & 96   & $8.60^{+0.24}_{-0.28}$& $7.72^{+0.18}_{-0.14}$& $0.59\pm0.06$\\[2pt]
\multicolumn{1}{c}{$9.2<$log(M$_\star$/M$_\odot$)$\leq10.54$} & 33   & $9.61^{+0.23}_{-0.24}$& $7.94^{+0.14}_{-0.12}$& $1.09\pm0.07$\\[2pt] \midrule
\end{tabular}

\tablefoot{\\ \tablefoottext{a}{The stellar masses are median values within each bin with the upper and lower errors representing the 84th and 16th percentile, respectively.}
\\ \tablefoottext{b}{The 12+log(O/H) measurements are based on the direct temperature method.}
\\\tablefoottext{c}{Obtained from dust-corrected H$\beta$ luminosities using the Table 2 from \cite{Theios2019_SFHa} that considers case B recombination, BPASSv2.2 100 $\rm M_\odot$, $Z_* = 0.004$ and a constant star-forming history and a minimum allowed age of 50 Myr. }}
\label{tab:mass_bins_Hg}
\end{table*}

\begin{table*}
\centering
\caption{Metallicity and SFR measurements in stellar mass bins for the H$\beta$ ($5.5<z<7$) sample.}
\resizebox{18cm}{!}{
\begin{tabular}{ccccccc}
\midrule
\multicolumn{1}{c}{Mass Bin}  & No.~ of Galaxies & log$\rm (M_{\star}/M_{\odot})$ \tablefootmark{a}& $\frac{\rm L_{[\ion{O}{iii}]\lambda5008}}{\rm L_{H\beta}}$& 12+log(O/H) (lower branch)\tablefootmark{b} & 12+log(O/H) (higher branch) & log$_{10}$(SFR / $\rm M_\odot\ yr^{-1}$)\tablefootmark{c}\\ \midrule
\multicolumn{1}{c}{$6.68\leq$log(M$_\star$/M$_\odot$)$<7.8$} & 159  & $7.36^{+0.29}_{-0.26}$& $ 6.47 \pm 0.26$ &$7.76^{+0.09}_{-0.08}$& $8.19^{+0.10}_{-0.10}$ & $0.24\pm0.02$  \\[2pt]
\multicolumn{1}{c}{$7.8<$log(M$_\star$/M$_\odot$)$<8.0$}     & 86   & $7.88^{+0.09}_{-0.09}$& $6.41\pm0.27$ &$7.75^{+0.08}_{-0.08}$& $8.21^{+0.09}_{-0.00}$&  $0.33\pm0.02$\\[2pt]
\multicolumn{1}{c}{$8.0<$log(M$_\star$/M$_\odot$)$<8.25$}     & 85  & $8.13^{+0.08}_{-0.09}$& $5.40\pm0.19$ &$7.51^{+0.04}_{-0.04}$& $8.54^{+0.07}_{-0.06}$&  $0.47\pm0.01$ \\[2pt]
\multicolumn{1}{c}{$8.25<$log(M$_\star$/M$_\odot$)$<8.5$}     & 91  & $8.35^{+0.09}_{-0.06}$& $6.97\pm0.26$ &$7.83^{+0.06}_{-0.06}$& $8.12^{+0.07}_{-0.07}$&  $0.41\pm0.02$\\[2pt]
\multicolumn{1}{c}{$8.5<$log(M$_\star$/M$_\odot$)$<8.75$}     & 100 & $8.63^{+0.08}_{-0.08}$ & $ 6.26\pm0.19$ &$7.72^{08}_{-0.06}$& $8.25^{+0.08}_{-0.09}$&  $0.53\pm0.01$\\[2pt]
\multicolumn{1}{c}{$8.75<$log(M$_\star$/M$_\odot$)$<9.0$}     & 71  & $8.84^{+0.11}_{-0.06}$& $6.25\pm0.18$ &$7.72^{+0.07}_{-0.06}$& $8.24^{+0.08}_{-0.08}$&  $0.62\pm0.01$\\[2pt]
\multicolumn{1}{c}{$9.0<$log(M$_\star$/M$_\odot$)$<9.5$}      & 79  & $9.21^{+0.18}_{-0.19}$& $6.77\pm0.21$ &$7.83^{+0.06}_{-0.07}$& $8.11^{+0.08}_{-0.07}$&  $0.77\pm0.01$\\[2pt]
\multicolumn{1}{c}{$9.5<$log(M$_\star$/M$_\odot$)$\leq11.1$} & 53   & $9.71^{+0.26}_{-0.17}$& $ 6.06 \pm 0.17$ &$7.66^{+0.06}_{-0.05}$& $8.32^{+0.06}_{-0.08}$& $0.83\pm0.01$  \\ \midrule
\end{tabular}

}
\label{tab:mass_bins_Hb}
\tablefoot{\\ \tablefoottext{a}{As in Table \ref{tab:mass_bins_Hg}, the stellar masses are median values within each bin with the upper and lower errors representing the 84th and 16th percentile, respectively.}
\\ \tablefoottext{b}{The 12+log(O/H) estimates are obtained from the R3 calibration from \cite{Chakraborty2025_MZR} and we show the results for both branches.}
\\ \tablefoottext{c}{The SFRs are based on the H$\beta$ line luminosity as in Table \ref{tab:mass_bins_Hg}.}
}
\end{table*}

\subsection{Measuring emission-line luminosity}
\subsubsection{Extracting 1D spectra}
\label{subsubsec: extracting_1D_spec}
We measure line luminosities based on optimally extracted 1D spectra \citep{1986PASP...98..609H,2023ApJ...950...67M}. We collapse the [\ion{O}{iii}]$\lambda 5008$ emission line over a wavelength range of $\pm 400$ km/s with respect to its wavelength ($5008.24$ \AA). We then perform non-linear least squares fitting of the spatial profile using the \texttt{lmfit} python package, finding that a Gaussian profile suffices. Subsequently, we collapse the 2D spectra by assuming the same spatial profile at all wavelengths.

The uncertainty on the 1D spectrum is obtained by propagating the uncertainties of the 2D spectrum that we estimated from bootstrapping.

\subsubsection{Emission line luminosity measurements}
We obtain measurements of the luminosities for [\ion{O}{iii}]$\lambda5008$, [\ion{O}{iii}]$\lambda4960$, H$\beta$, H$\gamma$ and [\ion{O}{iii}]$\lambda4364$ from our 1D continuum-subtracted stacks using the \texttt{lmfit} python package. We follow a similar methodology as presented in \cite{2023ApJ...950...67M}. We have the best SNR for the [\ion{O}{iii}]$\lambda\lambda 4960,5008$ emission lines which might aid in detecting the natural broadening of the emission lines that give it non-Gaussian wings (Lorentzian wings). However, in our case the extended shape in the wings of the lines are most likely due to the spatial-spectral degeneracy inherent to grism data (i.e. the spatial extent of the galaxies in the dispersion direction broadens the lines in the grism spectra). Therefore, for [\ion{O}{iii}]$\lambda5008$ and [\ion{O}{iii}]$\lambda4960$ we use a Voigt profile. We fix the standard deviation ($\sigma$) and the Gamma parameter to be the same for the [\ion{O}{iii}] doublet. We also fix the emission-line ratio between the doublet to be 2.98. \footnote{The Bayesian Information Criterion (BIC) values of the Voigt profiles for the [\ion{O}{iii}]$\lambda 4960, 5008$ doublet with the same standard deviation and Gamma parameter values and a fixed emission-line ratio are significantly lower than the BIC values for Gaussian profiles with the same standard deviation and fixed emission-line ratio. The absolute difference in the BIC values is larger than 10, with the BIC values for the Voigt profiles being lower. Therefore, a Voigt profile is preferred.} We fit a similar Voigt profile to the H$\beta$ line so that we trace the emission lines coming from the same gas. We fix the standard deviation, Gamma parameter and the amplitude ratio between the Gaussian and Lorentzian profiles to match the [\ion{O}{iii}$\lambda 5008$] line. The other emission lines are much fainter, therefore a Gaussian profile is used as a Voigt profile does not improve the reduced $\chi^2$ of the fits nor is there significant flux in the wings. The errors on the emission-line luminosities are propagated from the standard error and the covariance matrix obtained from \texttt{lmfit}. 

\subsection{Dust correction}
\label{subsec: dust_corr}
The luminosities of emission lines can be affected by dust attenuation, thus it is important to account for it to accurately estimate the gas-phase metallicity for galaxies. Although, dust attenuation correction for emission line ratios such as [OIII]$\lambda 5008/ \rm H \beta$ should negligible as the emission lines are close in wavelength. We calculate the intrinsic, dust-corrected luminosity for [\ion{O}{iii}]$\lambda 5008$, [\ion{O}{iii}]$\lambda 4364$ and H$\beta$ assuming the \cite{1989ApJ...345..245C} nebular attenuation curve in our H$\gamma$ sample stack,
\begin{equation}
     L_{\rm int}=L_{\rm obs}\cdot10^{0.4 {E(B-V)}\cdot k_{\lambda} },
\end{equation}
where
\begin{equation}
    E(B-V) = 0.95\cdot \tau,
\end{equation}
and
\begin{align*}
     \tau &= \log_{10}\left(\frac{0.473}{F_{\rm H\gamma}/F_{\rm H\beta}}\right)&{\rm for} &\ F_{\rm H\gamma}/F_{H\beta} \leq 0.473\,,\\
   \tau &= 0, &{\rm for} &\ F_{\rm H\gamma}/F_{\rm H\beta} > 0.473\,.
\end{align*}

$k_{\lambda}$ is the attenuation curve which is wavelength dependent and 0.473 is the intrinsic H$\gamma$/H$\beta$ line flux ratio that is appropriate for Case B recombination at $10^4$ K \citep{2006agna.book.....O}. We measure typical H$\gamma$/H$\beta$ values of $0.37\pm 0.04$ at average [\ion{O}{iii}] temperatures of $\sim 16\,600$~K. This indicates some level of dust attenuation, however, we find that the average dust-corrected [\ion{O}{iii}]$\lambda$5008/H$\beta$ (R3) and  [\ion{O}{iii}]$\lambda$5008/[\ion{O}{iii}]$\lambda$4364 ratios are only 0.01 dex and 0.03 dex lower than the average dust-uncorrected ratios, respectively.

\begin{figure}[h!]
    \includegraphics[width=9cm]{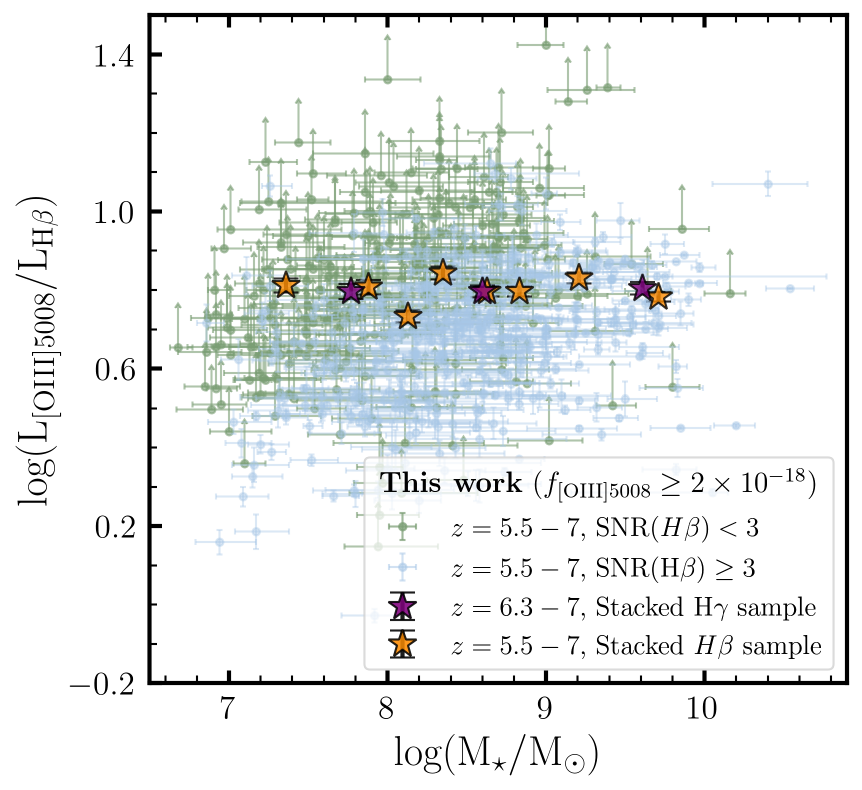}
      \caption{The R3 ratio as a function of stellar mass ratios for the entire dataset. Galaxies with SNR(H$\beta$)$\geq$3 are shown in blue and galaxies with SNR(H$\beta$)$<3$ are shown in green with lower limits. Stars show measurements in stacked spectra.}
         \label{fig: R3_ratios}
   \end{figure}

The $\rm [OIII]\lambda 5008/H\beta$ ratios for the H$\beta$ sample stacks are not dust-corrected as we do not always cover the H$\gamma$ line required to calculate Balmer decrements and the lines are close in wavelength range. In Fig. \ref{fig: R3_ratios}, we show the $\rm [OIII]\lambda 5008/H\beta$ ratios obtained for our stacked samples, along with ratios for individual galaxies in our entire dataset. For our stacks, we consider all H$\beta$ observations as focusing only on SNR(H$\beta$)$\geq$3 detections would bias our stacked values to lower ratio values in the low mass end. Applying a selection cut based on SNR(H$\beta$) would impose additional biases in our sample that impact our analysis of the mass-metallicity relation. The $\rm [OIII]\lambda 5008/H\beta$ ratio seems relatively invariant with stellar mass for the stacked samples. Additionally, the average ratios for both H$\gamma$ and H$\beta$ sample stacks are within 0.0009 dex of each other, which suggests that results obtained for the H$\gamma$ sample can be applied to the H$\beta$ sample (see Fig. \ref{fig: R3_ratios}).

\section{Metallicity Measurements}
\label{sec: measurements}

\subsection{Direct Method Metallicity measurements}
\label{subsec: Direct_Te}
We follow the prescription described in \cite{2006A&A...448..955I}, to estimate the gas phase metallicity using the direct-T$\rm_e$ method. For our H$\gamma$ sample, we detect the faint auroral line [\ion{O}{iii}]$\lambda 4364$ with S/N$>$3 in the stellar mass binned stacks. 
As the [\ion{O}{iii}]$\lambda4364$/[\ion{O}{iii}]$\lambda5008$ ratio is highly temperature dependent, we can use these detections to derive the electron temperature and the metallicities directly. We do not detect any [\ion{O}{ii}] emission lines that are required to calculate the electron density and the ionisation state of the gas as they are out of our observable spectral range. Therefore, we assume an electron density (n$_e$) of 300 cm$^{-3}$ as reported for similar low-metallicity, high-redshift galaxies in the literature \citep{2016ApJ...816...23S, Curti2024MZR, 2023ApJ...956..139I, 2023ApJ...950...67M, Sanders2024_MZR, Topping2025_eden}.
Recent work by \cite{Scholte2025_EXCELS} and \cite{Pollock2025_MZR}, probing the ISM for galaxies between $z\sim5.5-10$, suggests that T$\rm_e$ measurements and derived abundances are relatively insensitive to the electron density values n$_e \lesssim 10^4 $ cm$^{-3}$. We find that for an electron density between $ 100\rm \leq n_e (cm^{-3}) \leq 10^4 \rm $, our calculated abundances change by 0.02 dex at most. We also assume the O32 ratio -- log([\ion{O}{iii}$\lambda 5008$]/[\ion{O}{ii}]$\lambda \lambda 2727,29$) -- to follow a Gaussian distribution with mean of 0.98 and standard deviation of 0.36, as observed by \citet{2023A&A...677A.115C} for galaxies in the JADES survey between $5.5<z<9.5$. We find that their sample probes a similar dynamic range in $\rm M_{UV}$ and [\ion{O}{iii}]$\lambda5008$/H$\beta$ ratio as our entire sample. We find that the average [\ion{O}{iii}]$\lambda5008$/H$\beta$ ratio for our H$\gamma$ sample lies 0.06 dex above the average value for $z\sim6$ galaxies in \cite{2023A&A...677A.115C}. This can imply that the O32 ratios may be higher than assumed, which would lead to lower metallicities. To confirm this, we would require individual measurements covering both the [\ion{O}{iii}] and [\ion{O}{ii}] doublets. We use the T$_{\rm e}$[\ion{O}{iii}]-T$_{\rm e}$[\ion{O}{ii}] relation given for low gas-phase metallicity from \cite{2006A&A...448..955I}. With the dust corrected [\ion{O}{iii}]$\lambda5008$/[\ion{O}{iii}]$\lambda4364$ and the assumed O32 ratio and n$_e$ values we estimate the O$^{2+}$  and O$^{+}$ gas temperature using the Python package \texttt{PyNeb} \citep{2013ascl.soft04021L}.

We calculate the total oxygen abundance assuming that there is only O$^{+}$ and O$^{2+}$ present in the HII region. As we do not detect \ion{He}{ii} $\lambda$4686, a high ionisation energy line, in our stacked spectra, we do not add O$^{3+}$ ion abundance to our oxygen abundance \citep{2006A&A...448..955I} as O$^{3+}$ requires an even higher ionisation potential energy to be excited:  
\begin{equation}
    \frac{O}{H} = \frac{O^{+}}{H} +\frac{O^{2+}}{H} \; .
\end{equation}
  
\noindent The temperature of O$^{2+}$ gas ranges from $1.3\times10^4$ K to $2.1\times10^4$ K in our stacked sample, and the temperature of O$^{+}$ gas ranges from $1.3\times 10^4$ K to $1.6\times10^4$ K. The metallicities of our H$\gamma$ sample are presented in Table \ref{tab:mass_bins_Hg}. In Fig. \ref{fig: Z_HBeta}, we show that our metallicity values for the stacks of the H$\gamma$ sample agree with the R3 calibration curve given by \cite{Chakraborty2025_MZR}, \cite{Sanders2024_MZR} and \cite{Nakajima2022}, discussed in more detail in Sect. \ref{subsec: Z_Calib}. The observed stellar masses of galaxies in our H$\gamma$ sample span a range of $\rm log(M_\star/M_\odot)= 6.9-10.5$ with gas-phase metallicities ranging between $12 + \log_{10}(\rm O/H) = 7.6-8.1$ ($Z/Z_\odot = 0.07-0.24$).

The uncertainties in our calculated gas temperatures and metallicities were calculated by perturbing the observed emission line luminosity values by their measured errors. We have 1000 realisations of the perturbation and calculate the median and the 16th and 84th percentile values.

\subsection{Strong line calibrations measurements}
\label{subsec: Z_Calib}
\begin{figure}[h!]
   \centering
    \includegraphics[width=8cm]{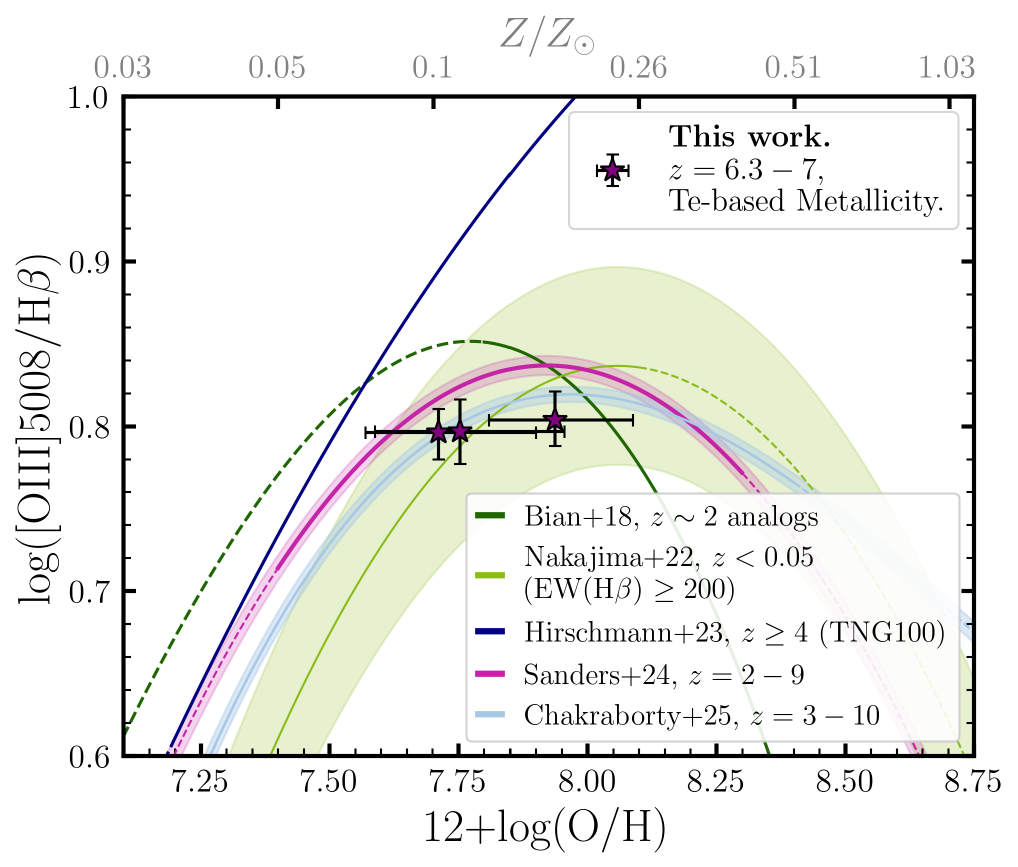}
      \caption{The relation between the gas-phase metallicity measured using the direct T$_e-$method and the R3 ratio. We compare our estimates with several strong-line metallicity calibrations from literature. We show the calibration by \cite{2018ApJ...859..175B}, which uses local analogues that occupy similar regions to $z\sim2$ star-forming galaxies on the BPT diagram, the calibration based on local galaxies with EW(H$\beta$)$\geq 200$ derived by \cite{Nakajima2022, 2023MNRAS.526.3504H} calibration for $z>4$ galaxies from the TNG100 simulation, and \cite{Sanders2024_MZR} and \cite{Chakraborty2025_MZR} calibrations for high-$z$ galaxies between $2<z<10$.}
      \label{fig: Z_HBeta}
   \end{figure}

For our H$\beta$ sample ($5.5 \leq z<7$), we do not cover the faint [\ion{O}{iii}]$\lambda 4364$ auroral line. Therefore, we estimate the metallicity based on the R3 ratio using the \cite{Chakraborty2025_MZR} and \cite{Sanders2024_MZR} calibration recently derived for high-redshift galaxies, and the \cite{Nakajima2022} calibrations derived for high H$\beta$ equivalent widths galaxies (EW(H$\beta)\geq200$\AA) in the local Universe (see Table \ref{tab:mass_bins_Hb}). Our choice for these calibrations is validated by our direct-T$\rm_e$ metallicity measurements for our H$\gamma$ sample (see Fig. \ref{fig: Z_HBeta}). To obtain the metallicities for the H$\beta$ sample, we solve the quadratic equation (cubic equation for \cite{Chakraborty2025_MZR}) for the R3-calibrations iteratively taking into account the observational error on the R3 ratio along with the errors on the calibration itself. We take the median of the solutions for the lower and higher branches to be the low and high metallicity estimate and the 16th and 84th percentile as the errors. As our spectra do not cover the [\ion{O}{ii}] emission lines, we cannot directly break the degeneracy between the ionization parameter and the metallicity that could both impact the [\ion{O}{iii}]$\lambda 5008$/H$\beta$ ratio. However, since our H$\gamma$ sample estimates fall either on the left branch or at the vertex of these relations, we assume that the galaxies in our H$\beta$ sample also fall on the left branch of the calibrations, which corresponds to high ionization and low metallicity. Our observations match best with the R3-calibration presented by \cite{Chakraborty2025_MZR}, which is based on direct-T$\rm _e$ metallicities for galaxies at $3<z<10$. We discuss the impact of using strong-line calibrations in the next section.\par

\section{The Mass - Metallicity Relation at $z\sim 6$}
\label{sec: MZR}

\begin{figure*}
   \centering
    \includegraphics[width=18cm]{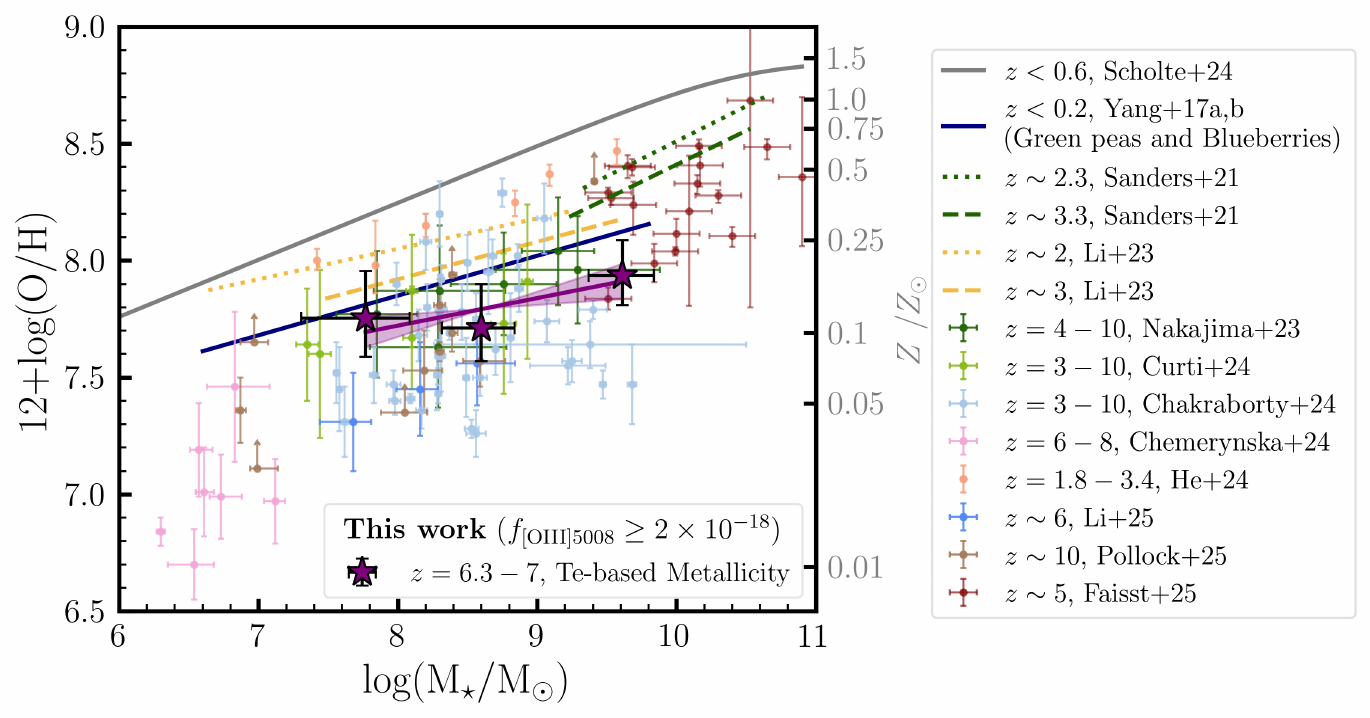}
      \caption{The mass-metallicity relation (MZR) from our stacked H$\gamma$ sample (purple stars, $6.3\leq z<7$). For the highest mass bin in the H$\beta$ sample we also show the metallicity estimate from the higher-metallicity branch of the \cite{Sanders2024_MZR} R3-calibration curve (unfilled black star). The metallicity estimates and median stellar masses for both the samples are reported in Table \ref{tab:mass_bins_Hg} and \ref{tab:mass_bins_Hb}. The errors on the stellar mass are calculated using the 16th and 84th percentile values in the stack. While our measurements are from stellar-mass binned and stacked NIRCam grism spectra, similar to \cite{2025arXiv250418616L}, measurements from \cite{Li23_z2_3} and \cite{He2024_MZR_NIRISS} are obtained using stacked spectra from NIRISS grism,   measurements from \cite{Chakraborty2025_MZR, Chemerynska2024} and \cite{Pollock2025_MZR} are from individual galaxies, values from \cite{Curti2024MZR} and \cite{2023ApJS..269...33N} are average metallicity measurements obtained from NIRSpec spectroscopy, and values from \cite{Faisst_ALPINE2025} are obtained from NIRSpec/IFU observations of high mass main-sequence galaxies with [\ion{C}{ii}]158 $\rm \mu m$ observations from ALMA.}
            
         \label{fig: MZR}
   \end{figure*}

Figure \ref{fig: MZR} shows the gas-phase oxygen abundances measured in our median stacked H$\gamma$ and H$\beta$ sample along with other literature measurements. We find a positive correlation between gas-phase metallicity and stellar mass for both of our subsets. We parametrise the MZR as
\begin{equation}\label{eq: MZR}
   \rm 12+log(O/H) = Z_{norm} + \gamma \cdot \left( log(M_\star/M_\odot) - 10 \right)\,,
\end{equation}
where for the H$\gamma$ sample we find that the best-fit normalisation is $\rm Z_{norm}=  7.96\pm0.10$, and the slope is $\gamma = 0.12\pm0.08$. The parameters are obtained though Orthogonal Distance Regression (ODR) using the \texttt{scipy.odr} package. Despite our lower sensitivity, our inferred metallicities agree with metallicity measurements, in similar redshift ranges, based on other surveys and instruments \citep[e.g. NIRSpec MSA,][]{Curti2024MZR, 2023ApJS..269...33N, Chakraborty2025_MZR}.

Our MZR lies systematically below the MZR in the local Universe and those at $z\sim2-3$. It lies 0.72 dex below the MZR given by \cite{Scholte2024_lowz} for galaxies in the local Universe.  At $z\sim 2$, it lies 0.55 dex, 0.35 dex and 0.61 dex below the MZR by \cite{Sanders2021MZR}, \cite{Li23_z2_3} and \cite{He2024_MZR_NIRISS}, respectively. It lies 0.45 dex, 0.28 dex and 0.63 dex below the MZR at $z\sim 3$ given by \cite{Sanders2021MZR}, \cite{Li23_z2_3} and and \cite{He2024_MZR_NIRISS}, respectively.

 \par

The normalisation of our H$\gamma$ sample MZR agree with previous indications of a redshift evolution in the MZR \citep[][and references therein]{2019A&ARv..27....3M, 2025arXiv250408933C}. We find that the slope of our MZR for the H$\gamma$ sample agrees well with slopes found for combined `Green Peas' and `Blueberries' sample \citep{Greenpeas_yang17, Blueberries_yang17} at $z<0.2$,  $z=2-3$ galaxies by \cite{Li23_z2_3} using NIRISS grism, and with sample from the JADES survey at $z=6-10$ \citep{Curti2024MZR}. In standard gas regulator models, invariance in slope of the MZR with increase in redshift may suggest invariance in physical mechanisms driving outflows in star-forming galaxies, whereas the lowering of normalisation at fixed stellar mass at higher redshifts indicates inflow of pristine gas \citep{Sanders2021MZR, 2023NatAs...7.1517H}. Table \ref{tab:MZR_tab} presents the slope and normalisation for the MZR for our study and other works. \par

\begin{figure}
   \centering
    \includegraphics[width=9cm]{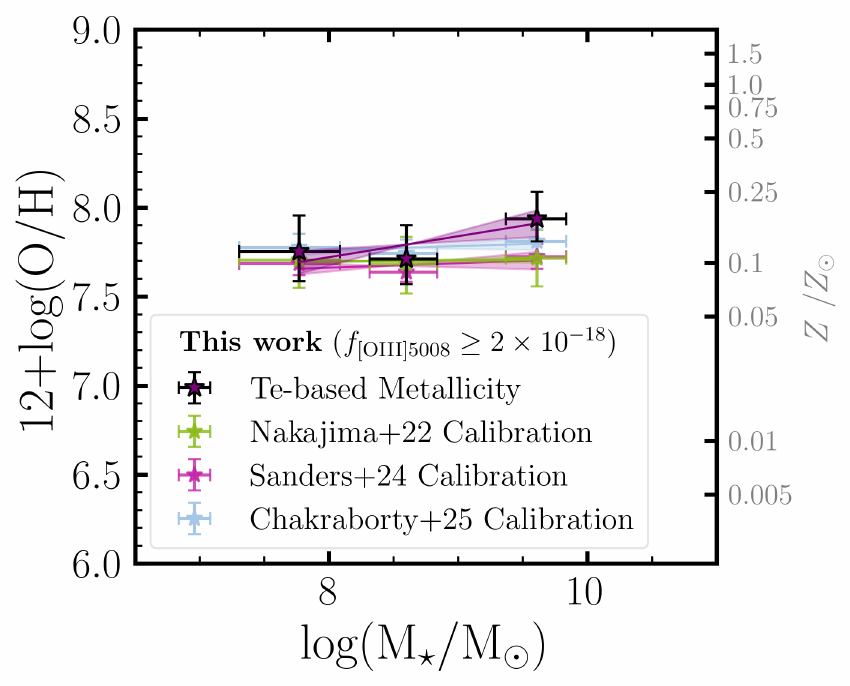}
      \caption{The mass-metallicity relation (MZR) inferred for our H$\gamma$ sample is shown using the direct-T$\rm_e$ method (purple stars), \cite{Chakraborty2025_MZR} R3-calibration (blue stars), \cite{Sanders2024_MZR} R3-calibration (magenta stars) and the \cite{Nakajima2022} R3-calibration for local galaxies with EW(H$\beta$)$>200 \AA$ (green stars). We use metallicity estimates from the lower branch of the R3-calibrations. The use of strong-line methods leads to an artificial flattening of the observed relation.}
         \label{fig: MZR_flat}
   \end{figure}

We also find that the slope of the MZR for the H$\beta$ sample is flatter than that of our H$\gamma$ sample. As described in previous sections, the metallicities for the H$\beta$ sample are obtained from the lower branch of the \cite{Chakraborty2025_MZR}, \cite{Sanders2024_MZR} and the \cite{Nakajima2022} calibrations. The slope of the MZRs obtained using  \cite{Chakraborty2025_MZR}, \cite{Sanders2024_MZR} and \cite{Nakajima2022} calibrations are 0.10, 0.066 and 0.023 respectively, which are $\sim 0.03,\ 0.06\ \rm and\ 0.1$ dex flatter than the slope of our direct-T$\rm_e$ estimates. Our R3 ratios for the H$\beta$ sample lie close to the vertex of the parabola which can make the MZR appear flatter. Indeed,  in Fig. \ref{fig: MZR_flat}, we show that calculating the metallicities of our H$\gamma$ sample stacks using the \cite{Chakraborty2025_MZR}, \cite{Sanders2024_MZR} and the \cite{Nakajima2022} calibrations significantly flattens the slope of the MZR as compared to the MZR from direct-T$\rm_e$ estimates. The general flatness of our MZR slope could also be due to the effects of our selection function, which we discuss in detail for the H$\gamma$ sample in Sect. \ref{sec: discussion}. 

While there are studies with which our MZR parameters agree with, the investigation of redshift evolution of the MZR is highly sensitive to systematic effects. For example, the best-fit parameters for our H$\gamma$ sample MZR compared to the MZR's from other studies in the same redshift range differ due to discrepancies in the methods used to obtain gas-phase metallicities, different stellar mass ranges probed, different selection functions and different parameters for SED fitting \citep{He2024_MZR_NIRISS}. Similarly, comparison of the MZR parameters for other redshift ranges is also affected by the same systematics. To understand the true evolution of the MZR, one should account for the selection effects for the various studies and estimate the metallicities using the same methodology. This is outside the scope of this work. However, we attempt to account for our observational selection and understand the nature of the intrinsic MZR in the Section ahead.\par

 \par

\section{Discussion}
\label{sec: discussion}

According to gas-regulator models, the slope of the MZR is governed by feedback mechanisms which drive gas out of galaxies \citep{PeeplesShankar2011,Sanders2021MZR, Curti2024MZR}. Such outflows are assumed to be driven by stellar feedback processes and they are expected to be more effective in removing metal-enriched gas from lower-mass galaxies as compared to the high-mass galaxies due to their shallower gravitational potential wells. The normalisation is controlled by the `effective yield' in the galaxy, which factors in the effect of nucleosynthetic stellar yields and gas flows. Finally, at fixed mass, the scatter along the MZR has been found to (anti-)correlate with the SFR of the galaxies. This has been observed as the FMR at $z\sim0$ \citep{Mannucci_FMR2010,Lopez_FMR2010,AandM_FMR2013, Curti_FMR2020}. The existence of the FMR at high-$z$ is still an unsolved question as many studies have shown that assuming the same FMR as at $z\sim0$ predicts higher metallicities than the ones they measure \citep{2014ApJ...792...75Z, Salim2015_FMR,2017ApJ...835...88K, Sanders2018_FMR, 2023NatAs...7.1517H, Curti2024MZR, Scholte2025_EXCELS, KCuestas2025_MZR}.

With recent results from JWST, new interpretations in understanding the MZR and FMR have arisen. JWST has shown that star-formation histories for high-$z$ galaxies usually are bursty \citep{Topping2022_burstySFH, Endsley2023_burstySFH,Looser2025_burstySFH}. The burst of star-formation may occur due to inflow of pristine gas, which dilutes the ISM, allowing for the galaxy to have a lower metallicity temporarily. However, there are other mechanisms through which the burst could occur (mergers, gravitational instabilities) which can cause an increase in the gas-phase metallicity \citep{Tacchella2023_burstySFH}. It has also been proposed that high-$z$ galaxies have not yet reached a stage of equilibrium between gas accretion, star formation, chemical evolution, and feedback processes \citep{Mannucci_FMR2010, Moller2013_FMR,2019A&ARv..27....3M, Curti2024MZR,Sarkar2025_MZR, Scholte2025_EXCELS,KCuestas2025_MZR}. Evidence for a non-equilibrium state of the ISM in star-forming galaxies has been observed through studies of the circumgalactic medium (CGM) of \ion{Mg}{ii} absorption systems along the sight-line of quasars at $z\sim6$ \citep{EIGER4_Bordoloi2024} and also in terms of turbulent gas kinematics of high-$z$ galaxies \citep{Danhaive2025_gas_kinematics}.\par 

In this section, we discuss the implications of our measurements and investigate the impact of our selection function using simulations and mock datasets. It is noteworthy that having a selection function based on $\rm H\beta$ fluxes, instead of $\rm [OIII]\lambda5008$ fluxes, would also impose selection effects on the observed MZR due to the possibly interrelated nature of the 3D relation between stellar mass, SFR and metallicity at these redshifts.

\subsection{Comparison to cosmological simulations}
\label{subsec: simulations}
The observed stellar masses of the galaxies in our H$\gamma$ sample span a range of $\rm log(M_\star/M_\odot)= 6.9-10.5$ with gas-phase metallicities ranging between $Z_{\rm gas} = 7.7-8.0$. We compare these measurements to those simulated in the SPHINX$^{20}$ \citep{Rosdahl2018_sphinx, 2023OJAp....6E..44K} and FLARES \citep{10.1093/mnras/staa3360,10.1093/mnras/staa3715} hydrodynamical simulations, for which [\ion{O}{iii}] and H$\beta$ luminosities are available. SPHINX$^{20}$ is a cosmological hydrodynamical radiative transfer simulation with a volume of $20^3$ cMpc$^3$. The simulation has been ran down to $z\approx4.5$. The FLARES simulations are a suite of 40 zoom-resimulations using the EAGLE model \citep{2015MNRAS.446..521S, 2015MNRAS.450.1937C}, which has been shown to reproduce galaxy properties over a wide range of redshifts including $z\approx0$. The zoom regions have been chosen to simulate a wide range of environments, from highly over-dense to under-dense. The FLARES simulations do not self-consistently model radiative transfer, but emission-line luminosities are added in post-processing based on Cloudy (version 17.03) modelling \citep{10.1093/mnras/staa3715}. For both simulations, we select galaxies between $z=5-7$ to compare to our dataset.  We use the stellar masses, dust attenuated emission-line luminosities and mass-weighted gas-phase metallicities from these simulations. In our comparison, we also include JAGUAR \citep{2018ApJS..236...33W} which is a  mock galaxy catalogue based on observationally driven empirical models of the evolution of galaxy properties.\par

\begin{figure}[h!]
   \centering
    \includegraphics[width=9cm]{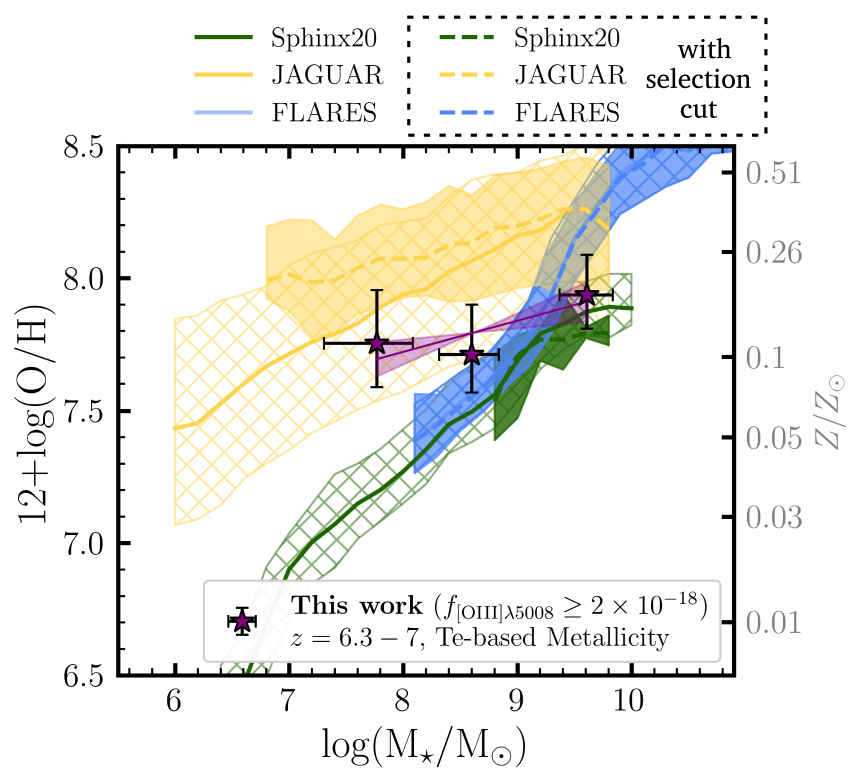}
      \caption{The mass-metallicity relation (MZR) for the SPHINX$^{20}$ \citep{2023OJAp....6E..44K} and FLARES  simulation \citep{10.1093/mnras/staa3360, 10.1093/mnras/staa3715}, the JAGUAR catalogue \citep{2018ApJS..236...33W}. We plot both, the intrinsic MZR and the MZR after mimicking our selection cut. We also show the MZR from our stacked H$\gamma$ (purple stars) sample. The MZR slopes and normalisations are listed in Table \ref{tab:MZR_tab}.}
         \label{fig: O3_sim}
   \end{figure}

In Fig. \ref{fig: O3_sim}, we show the MZR from FLARES, SPHINX$^{20}$ and JAGUAR at $z\sim6$. We use the mass-weighted gas-phase metallicity from the SPHINX$^{20}$ and FLARES simulation. \cite{2023ApJS..269...33N} make use of
stellar metallicities of young star particles ($< 10$ Myr) from FLARES for comparison to their observations, assuming that the stellar and gas-phase metallicities in the region of massive-star formation are comparable. We find that the slope ($\gamma$) and normalisation ($\rm Z_{norm}$) of the two MZRs from FLARES differ by 0.03 dex and 0.17 dex respectively, with the MZR from stellar metallicities of young star particles being slightly steeper. From the JAGUAR catalogue, we use the gas-phase metallicities which have been estimated using the \cite{Pettini_and_Pagel2004} N2-metallicity calibration.
We also plot our measurements based on the direct-$\rm T_e$ method. We find that simulations differ significantly among themselves. The slope and normalisation for the simulations, following Eq. \ref{eq: MZR}, have been noted in Table \ref{tab:MZR_tab}. The number of high-mass galaxies is limited in SPHINX$^{20}$ and JAGUAR by the simulated volume. The low mass cut-off determined by the resolution of the simulation, which is higher for SPHINX$^{20}$, therefore it includes lower mass galaxies. At face value, our MZR is flatter than the one in FLARES and SPHINX$^{20}$.\par

It is important to consider that our observational sample has been selected based on [\ion{O}{iii}] emission, which strength is sensitive to both star formation rate and metallicity, and may therefore yield a biased view of the MZR. In order to investigate the impact of our [\ion{O}{iii}]-flux limited galaxy sample on the observed MZR, we implement our selection cut ($f_{[\ion{O}{iii}\lambda5008]}\geq2\cdot10^{-18}\rm erg\ s^{-1} \ cm^{-2}$) on the [\ion{O}{iii}]$\lambda 5008$ fluxes in these simulated datasets. Before comparing the impact of the selection function on the observed MZR, we first compare our observed the [\ion{O}{iii}] luminosity as a function of stellar mass to that in simulations. We find that FLARES and JAGUAR galaxies have roughly similar relations between [\ion{O}{iii}] luminosity and stellar mass as our observed data (see Appendix \ref{Appendix: Simulations}). The SPHINX$^{20}$ simulation however shows systematically higher masses at fixed [\ion{O}{iii}] luminosity, which stems from the fact that SPHINX$^{20}$ has an elevated stellar mass - halo mass relation (i.e. galaxy formation is too efficient) and which therefore assigns too low SFRs (and consequently, [\ion{O}{iii}] luminosity) at fixed mass \citep{2023OJAp....6E..44K}. As a result, after applying our selection cut, the SPHINX$^{20}$ simulation only considers galaxies with $\rm log(M_\star/M_\odot) > 8.4$.\par 

For the MZR, we find that our selection function causes the observed slope to be flatter in JAGUAR and SPHINX$^{20}$ simulations. For JAGUAR, this is because our selection cut primarily removes low-metallicity galaxies at low masses. In SPHINX$^{20}$, along with the selection described before, it primarily removes massive galaxies with relatively high metallicities. Both effects effectively lead to a flattening. Our [\ion{O}{iii}] selection does not affect the observed MZR slope from FLARES. As discussed in the next Section, the effect of the [OIII] selection on the shape of the MZR depends on the correlation metallicity and the SFR at fixed mass. In FLARES, variations in metallicity are independent of SFR, such that the selection effect does not bias the observed shape of the MZR. The slopes for the [\ion{O}{iii}]-flux limited MZR are also summarised in Table \ref{tab:MZR_tab}.  We find a flatter MZR (slope, $\gamma = 0.12\pm0.08$) from our H$\gamma$ sample as compared to SPHINX$^{20}$ ($\gamma_{\rm sel} = 0.22$) and FLARES ( $\gamma_{\rm sel} = 0.44$) MZR even after mimicking the selection cut in the simulations. Therefore, we conclude that none of these hydrodynamical simulations is able to match our observed MZR, neither in slope nor in normalisation.\par 
We note that the $\rm L_{[OIII]}-M_\star$ relation from SPHINX$^{20}$ does not match our observations due to the low SFRs of SPHINX$^{20}$ galaxies (see Appendix \ref{Appendix: Simulations}). Since the gas-phase metallicity is inversely correlated to the sSFR in the simulation, our selection cut ends up selecting relatively high-metallicity galaxies and so the `observed' MZR also does not match. In SPHINX$^{20}$, we also miss out the highest mass and metallicity galaxies due to them being very dust-obscured and thus not meeting our [OIII] flux threshold. It is important to mention that we cannot account for missing out [OIII]$\lambda \lambda 4960,5008$ emitting galaxies in our observational data due to extreme dust-obscuration. This might contribute to the lack of a match between our observations and the simulations. There are no strong indications that we miss a large population of metal enriched, dusty galaxies. For example, \cite{Rowland2025_Rebels} observe the [OIII] doublet for dusty galaxies at $6<z<8$ with $\rm M_\star >10^{9} M_\odot$ from the REBELS ALMA program. Their sample is found to be more metal-rich than the bulk of the $z> 6$ population in the literature. However, we note that with their [OIII] luminosities, such galaxies would have all been selected if they were covered by our observations.

\subsection{Insights from modelling selection effects}
\label{subsec: toy_models}
Since the simulations do not match our observations and have varying response to our selection function, we investigate the effects of our [\ion{O}{iii}] flux-limited selection function on the observed mass - metallicity relation using simple toy models. In these models we explore the various dependencies between metallicity and star formation rate at fixed stellar mass. We start by sampling the stellar masses for galaxies using the galaxy stellar mass function (SMF) from \cite{Weibel_SMF2024} at $z=6$. Then, we assume the following linear relation between star formation rate and the stellar mass as has been found by multiple observational studies \citep[e.g.][]{Speagle2014, Popesso2023SFMS}:
\begin{equation}\label{eq: SFMS}
    \rm log_{10}(SFR) = MS_{norm} + MS_{slope}\times log_{10}\left(\frac{M_\star}{10^{10} M_\odot
}\right) + \mathcal{N}(0,\sigma_{SFR})\,,
\end{equation}
where the main sequence slope $\rm MS_{slope} = 1.0$ (motivated by the absence of strong evolution in the faint-end slope of the galaxy stellar mass functions, \citealt{Leja_2015_SMF}), the main sequence normalisation is $\rm MS_{norm} = 2.0$ (i.e. $\rm SFR = 100\ M_\odot/yr\ at\  log(M_\star/M_\odot) = 10$) and the intrinsic scatter along the relation is given by a normal distribution with standard deviation $\sigma_{SFR}=0.3$ (as found in \cite{DiCesare_SFMS}(Submitted to A$\&$A) for $\rm log(M_\star/M_\odot)=8$). Additionally, we consider the following 3D relation between gas-phase metallicity, stellar mass, and star formation rate:
\begin{equation}\label{eq: FMR eq}
    Z = Z_{\rm 0} + \rm M_{coeff} \times log(M_\star/M_\odot) + S_{coeff}\times log(SFR/M_\odot yr^{-1})
\end{equation}

\noindent where, $Z_{\rm 0}$ is the normalisation, $\rm M_{coeff}$ and $\rm S_{coeff}$ are the coefficients that determine the dependency between metallicity on stellar mass and SFR (note that $\rm M_{coeff}$ is not equal to the slope in the MZR due to the SFR dependence on mass). We then estimate H$\beta$ luminosities for our mock galaxies considering case B recombination and using the SFR-L$_{\rm H\alpha}$ relation from \cite{Theios2019_SFHa}, that assumes a Kroupa IMF with upper mass cut off of $100\rm M_\odot$ and slope -2.35 for $\rm M_\star>1M_\odot$ and, BPASSv2.2 model with constant SFH and stellar metallicity of $Z_\star =0.004$. Finally, we obtain [\ion{O}{iii}]$\lambda5008$ luminosities using the calibration between the R3 ratio and gas-phase metallicity from \cite{Sanders2024_MZR}.\par

Since we make various assumptions while generating the toy models, there are some caveats to keep in mind. Unlike our assumption, the scatter along the SFR-$\rm M_\star$ relation may be mass dependent, which would introduce a mass dependency in the scatter along the FMR. We also assume a linear conversion between SFR and L$_{\rm H\alpha}$ as prescribed by \cite{Theios2019_SFHa}, which has been assumed in literature extensively, but which may not be the most optimal in the early Universe \citep{Kramarenko_Sphinx_Ha}(Submitted tiA$\&$A). We assume the R3-calibrations from  \cite{Chakraborty2025_MZR}, without any scatter, to get [\ion{O}{iii}]$\lambda5008$ luminosities. It is plausible that for a given metallicity and H$\beta$ luminosity, there is scatter in this calibration that could be due to variations in the ionization parameter. Finally, we ignore the effects of dust attenuation motivated by the low attenuation found in Section $\ref{subsec: dust_corr}$.

\begin{figure*}
   \centering
   \begin{tabular}{ccc}
    \includegraphics[width=18cm]{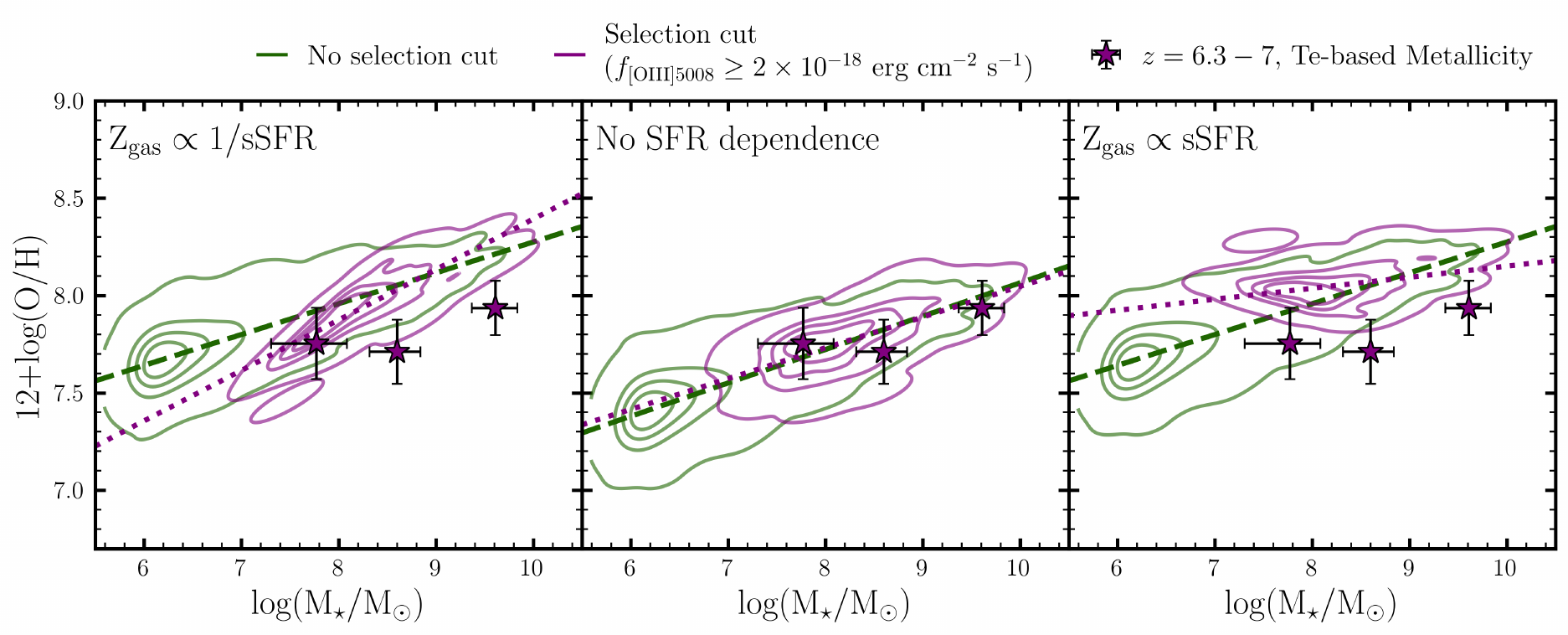}\\
     \includegraphics[width=18cm]{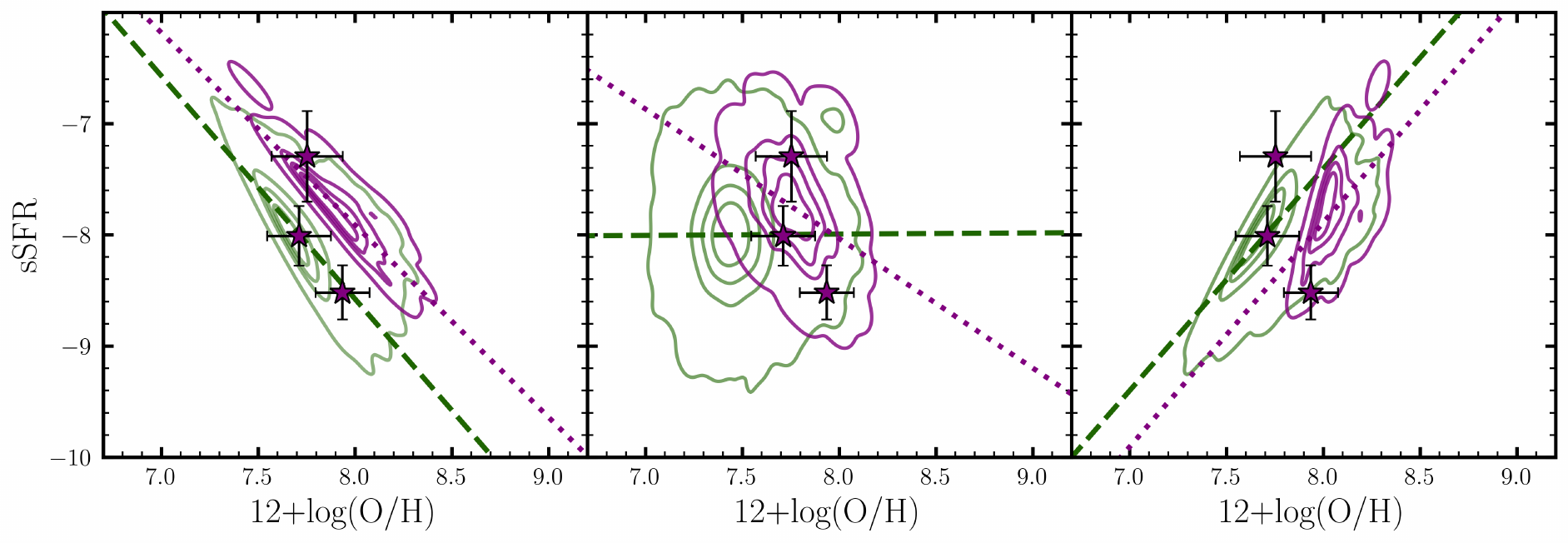}
    \end{tabular}
      \caption{(Top) The MZR for the three cases in our toy model. We also plot our direct-T$_e$ metallicity estimates. (Bottom) Specific star formation rate (sSFR) as a function of gas-phase metallicity. We show sSFR vs. gas-phase metallicity for our H$\gamma$ stellar mass binned stacks. The errorbars on the sSFR are calculated using the 16th and 84th percentile in the stacks. The top left and bottom left panel shows Case 1 (`normal FMR'). Our selection cut makes the MZR slope steeper . The top middle and bottom middle panel shows Case 2 (`no FMR'). Our selection cut does not affect the MZR slope. The top right and bottom middle panel shows Case 3 (`inverted FMR'). Our selection cut makes the MZR slope appear flatter. In all three cases, we find a negative correlation between sSFR and 12+log(O/H) upon implementing our selection cut.}
         \label{fig: toy_model}
   \end{figure*}

We generate mock datasets for three cases: 
\begin{enumerate}
    \item {\bf Normal FMR case ($\rm S_{coeff}<0$):} There exists a 3D relation between gas-phase metallicity, stellar mass and star formation rate, such that gas-phase metallicity is inversely proportional to the specific star-formation rate. This is seen in galaxies in the local Universe and understood in the context of gas regulation models.
    
    \item  {\bf No FMR case ($\rm S_{coeff}=0$)}: In this scenario, there is no 3D relation between gas-phase metallicity, stellar mass and star formation rate. The gas-phase metallicity only depends on the stellar mass (MZR) and the scatter in the MZR is unrelated to SFR. This is the case in the FLARES simulation.
    
    \item {\bf Inverted FMR case ($\rm S_{coeff}>0$):} There exists a 3D relation between gas-phase metallicity, stellar mass and star formation rate, but here gas-phase metallicity is directly proportional to the specific star-formation rate. This is seen on resolved (100 pc) scales in local galaxies \citep{Wang_and_lilly_2021} and suggests delayed or inefficient feedback.  
\end{enumerate}
The parameter choices for the three cases are given in Table \ref{tab:toy_models}. We motivate these three cases as they have a distinct impact on how our selection function influences the observed shape of the MZR. According to `gas-regulator models', chemical enrichment in galaxies is set by equilibrium between gas accretion, star-formation and outflows \citep{ 2008MNRAS.385.2181F, PeeplesShankar2011, 2012MNRAS.421...98D, 2013ApJ...772..119L}. Under such equilibrium, inflow of low-metallicity gas dilutes the ISM and triggers star formation. The star formation consumes the metal-poor gas and enriches the ISM, thus increasing the metallicity until new gas is accreted. This gives rise to the anti-correlated behaviour of the gas-phase metallicity and sSFR, which has been observed by multiple observational studies in the local Universe as well in the early Universe \citep[e.g.][]{Mannucci_FMR2010,Lopez_FMR2010,Sanders2018_FMR}. This motivates the first case, `normal' FMR case, in our toy models.  The parameter values for this case are taken from \cite{AandM_FMR2013}, considering their 2D-projection of the FMR. The scatter on $Z_{\rm gas}$ comes from the scatter in the star-forming main sequence and strength of $\rm S_{coeff}$ (Eq. \ref{eq: SFMS}).\par
However, it is unclear whether the `normal' FMR evolves out to high redshift as multiple studies have also found discrepancies in their observations \citep{2014ApJ...792...75Z, 2023NatAs...7.1517H, Curti2024MZR, Scholte2025_EXCELS, 2025MNRAS.540.2176C}. In a recent study, \cite{Laseter_lowz_FMR2025} find no evidence for the existence of the `normal' FMR  at $z\sim0$ for the low-$\rm M_\star$ regime indicating deviation from the equilibrium conditions as seen at the higher-$\rm M_\star$ regime. While the gas-regulator models explain the existence of FMR, there are many caveats. For example, supernovae outflows are the only feedback assumed in the models. Additionally, these models assume time-invariant star-forming efficiencies, which may not be true for galaxies with bursty star-formation histories. Importantly, the 3D relation is not just set by the strengths of processes such as the mass inflow rate, star formation rate, mass loading factor, ISM metallicity, effective yields and star forming efficiency, but also on the times scales of their variations \citep{Torrey2018_FMR, Wang_and_lilly_2021}. \cite{Torrey2018_FMR} show that if the time-scales for variations in the SFR and gas-phase metallicity are within a factor of $\sim 1.5$ of each other in the TNG100 simulation, then the `normal FMR' case should hold. However, if either the SFR or metallicities evolves much faster than the other it can either wash out any correlation between them or cause a positive correlation. This motivates our second case (no FMR) and third case (inverted FMR). Processes like bursty star formation can significantly impact the variability of SFR while not impacting the metallicity much. Multiple studies have shown evidence for bursty star formation for high-redshift galaxies \citep{Topping2022_burstySFH, Endsley2023_burstySFH,Looser2025_burstySFH}. Additionally, variability in processes such as temporary increase in the SFE due to gravitational instabilities or, increase in outflows with the same metallicity as the ISM can allow for a positive correlation between sSFR and gas-phase metallicity \citep{Torrey2018_FMR, Wang_and_lilly_2021}. Indications of an `inverted' FMR has also been seen in the THESAN-ZOOM simulation for galaxies with $\rm log(M_\star/M_\odot)\lesssim9$ due to regular inflow of low metallicity gas \citep{McClymont2025_ThesanZoom}. The parameters for the second case (No FMR) have been taken from the MZR presented in \cite{Curti2024MZR}. We assume a scatter of 0.1 dex on the MZR. 
For the third case, we obtain the parameters for Eq. \ref{eq: FMR eq} by matching the shape of the intrinsic MZR to the one obtained considering the `Normal MZR' case.  We keep the $\rm S_{coeff}$ to be the same of the magnitude as for the normal FMR case and only change the sign. In this case, we find $\rm M_{coeff}<0$. As per Eq. \ref{eq: FMR eq}, this does not imply a negative correlation between metallicity and stellar mass as SFR is also a function of stellar mass (see Appendix \ref{Appendix: FMR_eq_model}). However, this does point toward processes such as bursty star formation and inflow of pristine gas to be more dominant as compared to the stellar mass at high redshift. \par

In the top panel of Fig. \ref{fig: toy_model}, we show the distribution of simulated galaxies on the mass - metallicity plane for the three cases mentioned above. We apply our selection cut on [\ion{O}{iii}]$\lambda 5008$ luminosity. We then compare how the MZR of the selected sample, i.e. `observable' MZR, compares to the intrinsic one. The selection cut affects the three cases differently. For case 1 (normal FMR), the observable MZR is steeper than the intrinsic MZR. In this case, $Z_{\rm gas} \propto \rm 1/sSFR$, therefore an [\ion{O}{iii}] flux cut selects galaxies with lower metallicity.  This makes the slope of the observable MZR steeper.  For case 2 (No FMR), there is negligible change in the intrinsic MZR and the observable MZR slopes. The small change in the slope can be attributed to the scatter added to the MZR. For case 3 (inverted FMR), the observable MZR is flatter than the intrinsic MZR. In this case, $Z_{\rm gas} \propto \rm sSFR$, thus [\ion{O}{iii}] flux scales with SFR, and we end up selecting galaxies with higher metallicities with an [\ion{O}{iii}] flux cut. This makes the slope of the observable MZR flatter.\par

With the current data available, it is not possible to determine which scenario our observations follow as there are multiple degeneracies in the free parameters while evaluating the metallicities using Eq. \ref{eq: FMR eq} (we discuss these degeneracies in Appendix \ref{Appendix: FMR_eq_model}). We attempt to test which scenario explains our observations by studying the sSFR as a function of the gas-phase metallicity as shown in the bottom panel of Fig. \ref{fig: toy_model}. 
We see a negative correlation between the sSFR and gas-phase metallicity for both the `intrinsic' and `observable' distribution in case 1. For case 2, there is no correlation between the sSFR and gas-phase metallicity for the intrinsic distribution, however, on application of our selection cut we find a negative correlation. In case 3, both the intrinsic and observable distributions have a positive correlation. We find a negative correlation in our observations, the slope of which is best represented by the `No FMR' case. This suggests an absent or weak ($\rm S_{coeff} =0\ or\  |S_{coeff}|<<1\ but \neq 0$ ) FMR at $z\sim6$ which may allude to the galaxies at these redshifts being out of equilibrium, as has been suggested previously for high-redshift galaxies \citep{2019A&ARv..27....3M, EIGER4_Bordoloi2024,Danhaive2025_gas_kinematics}. In addition to the MZR parameters from \cite{Curti2024MZR} for case 2, we also explore the impact of using parameters from other MZRs in the literature \citep{2023NatAs...7.1517H, Sarkar2025_MZR} (see Appendix \ref{Appendix: MZR_models}). We find that our observations generally agree with the MZR models in the sSFR vs. gas-phase metallicity plane. If this is the true underlying scenario, then our intrinsic slope is as flat as our observed slope, which points towards either rapid metal enrichment in low-mass galaxies or lower metallicities due to pristine gas inflows in higher mass galaxies. Previous observations of high-$z$ galaxies are not in favour of lower metallicities for high stellar mass galaxies \citep{Rowland2025_Rebels}, suggesting the former explanation. Observations of CGM galaxies at $z\sim6$ also indicate that galaxies in the early Universe were efficient in chemically enriching their surroundings \citep{EIGER4_Bordoloi2024}. \par

In this case, it is important to note that a good fit between the model and the data does not necessarily indicate that the intrinsic trend is represented by the model. Fig. \ref{fig: toy_model} shows that the scatter in the sSFR - metallicity relation varies across models.  Additionally, Fig. \ref{fig: MZR_toy_models} (see Appendix \ref{Appendix: MZR_models}) shows that scatter varies in the sSFR - metallicity plane even for the same model (case 2, $\rm S_{coeff} =0$) with different parameters. Therefore to fit the parameters and distinguish between the models, we need to fit the scatter. Due to the stacked measurements of the metallicity, we cannot measure the scatter in the sSFR - metallicity relation and therefore cannot fully constrain the SFR dependence. To do this we require observations with robust metallicity measurements for a statistically large sample of individual galaxies with a well defined selection function, as this crucially impacts the observed relations.

\begin{table}[]
\caption{Parameter choices for the toy models investigated in Sect. $\ref{subsec: toy_models}$ (Eq. $\ref{eq: FMR eq}$).}
\label{tab:toy_models}
\resizebox{\columnwidth}{!}{
\begin{tabular}{ccccc}
\midrule
Models  & $Z_{\rm norm}$ & $\rm M_{coeff}$ & $\rm S_{coeff}$ & \multicolumn{1}{c}{Comments} \\ \midrule
Case 1:  $Z \propto 1/$sSFR & 4.2  & 0.47 & $-0.3102$ & \makecell[l]{Values adopted from \\\cite{AandM_FMR2013} \\ FMR}                                 \\ \midrule
Case 2:  No SFR dependence           & 6.39 & 0.17 & -       & \makecell[l]{Values adopted from \\ \cite{Curti2024MZR}.\\ We consider a \\0.1 dex  scatter \\along the relation }\\ \midrule

Case 3:  $Z \propto$ sSFR & 9.15         & $-0.15$            &     0.3102       &                              \\ \midrule
\end{tabular}
}
\end{table}

\section{Summary}
\label{sec: summary}
In this work, we study and characterise the MZR and the 3D correlation among stellar mass, gas-phase metallicity, and SFR, dubbed as the fundamental metallicity relation (FMR), of [\ion{O}{iii}] selected galaxies with redshifts between $5 < z <7$. We obtain the spectra for these galaxies using deep NIRCam WFSS surveys, EIGER \citep{2023ApJ...950...66K}, COLA1 \citep{2024A&A...689A..44T} and ALT \citep{2024arXiv241001874N}. Grism spectra have a well-understood flux-limited selection function and are not subject to slit losses. To make the selection uniform across the data from the three surveys uniform, we make a conservative [\ion{O}{iii}] flux cut, $f_{\ion{O}{iii}\lambda5008}\geq 2\times10^{-18}\rm  erg\ s^{-2}\ cm^{-2}$. Since grism spectra are less sensitive and have a smaller wavelength coverage compared to NIRSpec spectra, we perform stacking in stellar mass bins to detect faint emission lines such as [\ion{O}{iii}]$\lambda4364$ and H$\gamma$ with a $\rm S/N>3$ (see Fig. \ref{fig: Hgamma_stacked_spec}). This allows us to derive Direct-T$\rm_e$ based metallicities for our analysis. We test the impact of our selection function using hydrodynamical simulations such as FLARES \citep{10.1093/mnras/staa3360,10.1093/mnras/staa3715} and SPHINX$^{20}$ \citep{Rosdahl2018_sphinx, 2023OJAp....6E..44K} and, the JAGUAR catalogue \citep{2018ApJS..236...33W}. We further test our selection function using simple toy models for three scenarios: (1) Normal FMR ($Z_{\rm gas} \propto \rm 1/sSFR$), (2) No FMR (No SFR dependence) and (3) Inverted FMR ($Z_{\rm gas} \propto \rm sSFR$).  Our main results are:

\begin{enumerate}

    \item We cover H$\gamma$ and [\ion{O}{iii}]$\lambda4363$ for a sample of 270 galaxies (i.e. the H$\gamma$ sample). We detect H$\gamma$, [\ion{O}{iii}]$\lambda4363$, H$\beta$ and [\ion{O}{iii}]$\lambda4960,5008$ when stacking in bins of stellar mass, which allows us to measure direct-T$\rm_e$ metallicities in the stacks. The observed stellar masses for this sample span a range of $\rm log(M_\star/M_\odot)= 6.9-10.5$ with the gas-phase metallicities ranging in $Z_{\rm gas} = 7.7-8.0$ ($0.11-0.22 Z_\odot$) [Sect. \ref{subsec: Direct_Te}, Fig. \ref{fig: MZR}, Table \ref{tab:mass_bins_Hg}].

    \item With the measured direct-T$\rm_e$ metallicities we test the R3 calibration curve for high-$z$ galaxies. We find that our direct-T$\rm_e$ measurements are slightly offset to higher values of the R3 ratio at fixed metallicity as compared to the R3 calibration curves from \cite{Sanders2024_MZR} and \cite{Nakajima2022}. However, within the observational errors and the errors on the calibration itself, our values agree with these curves [Sect. \ref{subsec: Z_Calib}, Fig. \ref{fig: Z_HBeta}]. 

    \item We obtain a linear MZR relation with normalisation $\rm Z_{norm}=  7,96\pm0.08$, and slope $\gamma = 0.12\pm0.08$ for the direct-T$\rm_e$ measurements from the H$\gamma$ sample [Sect. \ref{sec: MZR}, Eq. \ref{eq: MZR}]. The MZR lies systematically below the MZR measured for galaxies in the local Universe \citep{Scholte2024_lowz} and at $z\sim2-3$ \citep{Sanders2021MZR, Li23_z2_3}. The slope of our MZR agrees well with those found for analogues at lower-redshift \citep{Greenpeas_yang17, Blueberries_yang17} and with other $z=6-10$ measurements \citep{Curti2024MZR} [Sect. \ref{sec: MZR}, Fig. \ref{fig: MZR}, Table \ref{tab:MZR_tab}].

    \item  We find that our measurements do not match the slope and normalisation of the MZR of galaxies in the FLARES and SPHINX$^{20}$ simulations and the JAGUAR model. The observed slope is significantly shallower, which is robust to our selection effects. Selection effects impact simulations non-uniformly. In SPHINX$^{20}$ and JAGUAR, on applying the selection cut, the MZR becomes flatter. For FLARES, there is no change in the MZR upon application of the selection cut [Sect. \ref{subsec: simulations}, Fig. \ref{fig: O3_sim}, Table \ref{tab:MZR_tab}]. 

    \item  Through simple toy models, we show that the effect of our [\ion{O}{iii}] selection cut on the MZR depends on the relation between gas-phase metallicity and sSFR of the galaxies. For the case of a normal FMR, the observable MZR (after the selection cut) is steeper than the intrinsic MZR.  For the case of no FMR, i.e. no SFR dependence, there is negligible change in the intrinsic MZR and the observable MZR slopes. For the case of an inverted FMR, the observable MZR is flatter than the intrinsic MZR [Sect. \ref{subsec: toy_models}, Fig. \ref{fig: toy_model}, Table \ref{tab:toy_models}]. Our data agrees well with the `no FMR' case. This suggests an absent or weak FMR at $z\sim 6$ which is similar to what we find in FLARES. An absent or weak FMR points towards the possibility that the galaxies at these redshifts are not yet in equilibrium. A `no FMR' case would imply that the intrinsic slope of the MZR is as flat as the observed one. Such flatness may be due to rapid metal enrichment [Sect. \ref{subsec: toy_models}, Fig. \ref{fig: toy_model}]. However, since our data points are obtained through stacking they are not enough to constrain the scatter in the three cases and therefore, distinguish cleanly among them. 
\end{enumerate}

This work highlights the importance of accounting for selection functions in order understand the physics underlying scaling relations such as the  MZR and the FMR. In order to understand the baryon cycle in depth at high redshifts, future observations with robust metallicity measurements for a statistically large well-defined samples of galaxies.

\begin{acknowledgements}

We thank the anonymous referee for the insightful comments that helped improving this paper. This work is based on observations made with the NASA/ESA/CSA James Webb Space Telescope.  The data were obtained from the Mikulski Archive for Space Telescopes at the Space Telescope Science Institute, which is operated by the Associations of Universities for Research in Astronomy, Inc., under NASA contract NAS 5-03127 for JWST.  These observations were taken under programs \# 1243, \# 1933 and \# 3516.

Funded by the European Union (ERC, AGENTS,  101076224). Views and opinions expressed are however those of the author(s) only and do not necessarily reflect those of the European Union or the European Research Council. Neither the European Union nor the granting authority can be held responsible for them. 

GK acknowledges support from the Foundation MERAC. APV acknowledge support from the Sussex Astronomy Centre STFC Consolidated Grant (ST/X001040/1). 

\end{acknowledgements}

\bibliographystyle{aa}
\bibliography{ref}

\begin{appendix} 

\section{Metallicity estimates for $\rm H \beta$ sample using R3-calibrations}
\label{Appendix: R_calibs_MZR}
We report the gas-phase metallicity estimates for our stellar mass bin stacked $\rm H\beta$ sample using the R3-calibrations presented in \cite{Nakajima2022} and \cite{Sanders2024_MZR} in Table \ref{tab:MZR_r3_calibs}. We use R3-calibration presented for local galaxies with $\rm EW{H\beta}\geq 200$  from \cite{Nakajima2022}. The gas-phase metallicities have been estimated using the same methodology as described in Sect. \ref{subsec: Z_Calib}.

\begin{table*}[h]
\caption{Gas-phase metallicity estimates in stellar mass bins for the H$\beta$ ($5.5<z<7$) sample using R3 calibration.}
\centering

\begin{tabular}{lccccc}
\midrule
\multicolumn{1}{c}{Mass Bin} &  log$\rm (M_{\star}/M_{\odot})$ &  \multicolumn{2}{c}{\cite{Sanders2024_MZR}} &  \multicolumn{2}{c}{\cite{Nakajima2022}} \\
\multicolumn{1}{c}{} &   & \makecell[l]{ $\rm 12+log(O/H)$ \\(lower branch)} & \makecell[l]{ $\rm 12+log(O/H)$ \\(higher branch)} &  \makecell[l]{ $\rm 12+log(O/H)$ \\(lower branch)} & \makecell[l]{ $\rm 12+log(O/H)$ \\(higher branch)} \\ \midrule
\multicolumn{1}{c}{$6.68\leq$log(M$_\star$/M$_\odot$)$<7.8$} &  $7.36^{+0.29}_{-0.26}$ & $7.67^{+0.09}_{-0.07}$ &   $8.17^{+0.07}_{-0.09}$ &  $7.72^{+0.16}_{-0.18}$ &  $8.40^{+0.18}_{-0.16}$ \\
$7.8<$log(M$_\star$/M$_\odot$)$<8.0$ &  $7.88^{+0.09}_{-0.09}$ &  $7.66^{+0.08}_{-0.07}$ &  $8.18^{+0.07}_{-0.08}$ &  $7.68^{+0.19}_{-0.14}$ &  $8.44^{+0.14}_{-0.19}$ \\
$8.0<$log(M$_\star$/M$_\odot$)$<8.25$ &  $8.13^{+0.08}_{-0.09}$ &  $7.44^{+0.04}_{-0.04}$ &  $8.40^{+0.04}_{-0.04}$ &  $7.60^{+0.16}_{-0.18}$ &  $8.52^{+0.18}_{-0.16}$ \\
\multicolumn{1}{c}{$8.25<$log(M$_\star$/M$_\odot$)$<8.5$} &  $8.35^{+0.09}_{-0.06}$ &  $7.77^{+0.08}_{-0.07}$ &  $8.07^{+0.07}_{-0.08}$ &  $7.74^{+0.15}_{-0.16}$ &  $8.38^{+0.16}_{-0.15}$ \\
\multicolumn{1}{c}{$8.5<$log(M$_\star$/M$_\odot$)$<8.75$} &  $8.63^{+0.08}_{-0.08}$ &  $7.62^{+0.06}_{-0.05}$ &
  $8.22^{+0.05}_{-0.06}$ &  $7.68^{+0.16}_{-0.17}$ &  $8.44^{+0.17}_{-0.16}$ \\
$8.75<$log(M$_\star$/M$_\odot$)$<9.0$ &  $8.84^{+0.11}_{-0.06}$ &  $7.62^{+0.06}_{-0.05}$ &  $8.22^{+0.05}_{-0.06}$ &  $7.70^{+0.15}_{-0.19}$ &  $8.42^{+0.19}_{-0.15}$ \\
$9.0<$log(M$_\star$/M$_\odot$)$<9.5$ &  $9.21^{+0.18}_{-0.19}$ &  $7.75^{+0.07}_{-0.07}$ &  $8.09^{+0.07}_{-0.07}$ &  $7.75^{+0.13}_{-0.17}$ &  $8.37^{+0.17}_{-0.13}$ \\
  $9.5<$log(M$_\star$/M$_\odot$)$\leq11.1$ &  $9.71^{+0.26}_{-0.17}$ &  $7.57^{+0.05}_{-0.04}$ &  $8.27^{+0.04}_{-0.05}$ &  $7.66^{+0.17}_{-0.17}$ &  $8.45^{+0.17}_{-0.17}$ \\ \midrule
\end{tabular}

\label{tab:MZR_r3_calibs}
\end{table*}

\section{MZR parameters for observations and simulations}
We report the slope and normalisation for the MZR for our direct-T$\rm_e$ measurements along with the observational samples and simulations, following Eq. \ref{eq: MZR}. For the simulations, we present the MZR parameters for a sample with the same selection cut as our observations (i.e. an [\ion{O}{iii}]-flux cut) and for a sample without any selection cut.

\begin{table*}[]
\caption{The slope of the mass-metallicity relation in various observational samples and in simulations.}
\label{tab:MZR_tab}
\resizebox{\textwidth}{!}{
\begin{tabular}{llllll}
\toprule
\multicolumn{1}{c}{Study}  &\multicolumn{1}{c}{Redshift}& $\rm [log(M_\star/M_\odot$)] & \multicolumn{1}{c}{$\gamma$}  & \multicolumn{1}{c}{Z$_{\rm norm}$} & \multicolumn{1}{c}{Description}\\ \midrule
This work (direct-Te)  & 6.3-7   & [6.89, 10.54]&     $0.12\pm0.08$          &   $7.96\pm0.10$  &   \makecell[l]{$f_{[\ion{O}{iii}]\lambda5008}$ selected galaxies \\from NIRCam/WFSS  (EIGER, \\COLA1 and ALT surveys)}   \\ \hline \midrule
 
  \makecell[l]{Greenpeas \citep{Greenpeas_yang17} \\and Blueberries \citep{Blueberries_yang17}        } &  $z<0.2$    &  [6.6,9.84]& $0.17\pm0.01$   & $8.19\pm0.03$ &  \makecell[l]{Analogues for high-$z$ galaxies \\ because of their compact size,\\ high equivalent width emission \\lines,  high ionisation parameters \\and low metallicities.}  \\ \midrule

\cite{Scholte2024_lowz}\tablefootmark{a} & $<0.6$ & [6.4, 11.05]  & $0.23\pm 0.003$ & $8.70 \pm0.003$   & \makecell[l]{Sample from the DESI BGS \\validation and Year 1 Main \\Survey Data \citep{DESI_2024_Hahn}} \\ \midrule

\cite{Sanders2021MZR}         & $\sim0.08$    & [8.5, 11.5]& $0.28\pm0.01$   & $8.77\pm0.01$   & \makecell[l]{$z\sim0$ sample from SDSS \\ (\citealt{2000AJ....120.1579Y},\\ \citealt{ AandM_FMR2013}).} \\
                      & 2.3     & [9.3, 10.7] & $0.30\pm0.02$   & $8.51\pm0.02$  & \makecell[l]{\\$z\sim 2-3$ star-forming \\galaxies from MOSDEF \\survey \citep{Kreik_MOSDEF2015}.} \\
                      & 3.3     & [9.2, 10.6] & $0.29\pm0.02$   & $8.41\pm0.03$  &  \\ \midrule
\cite{Li23_z2_3}         & $\sim2$     & [6.52, 9.45] & $0.13\pm0.04$   & $8.31\pm0.04$  &  \makecell[l]{JWST/NIRISS slitless grism \\spectroscopic observations \\in the Abell 2744 and SMACS \\J0723-3732 fields.} \\
                      & $\sim3$     & [7.39, 9.68] & $0.16\pm0.03$   & $8.24\pm0.03$ & \makecell[l]{Uses the O32 calibration from \\ \cite{2018ApJ...859..175B}.}\\ \midrule
\cite{He2024_MZR_NIRISS}        & $1.90$     & [6.9, 9.9) & $0.223\pm0.017$   & $8.569\pm0.036$  &  \makecell[l]{JWST/NIRISS slitless grism \\spectroscopic observations \\ in the Abell 2744 field.}\\
                      & $2.88$     & [7.1, 10.0) & $0.294\pm0.010$   & $8.596\pm0.024$ & \makecell[l]{Uses calibrations from \\ \cite{2018ApJ...859..175B}.}\\ \midrule
\cite{2023ApJS..269...33N}         & $4-10$     & [7.39, 9.48] & $0.25\pm0.03$   & $8.24\pm0.05$ & \makecell[l]{ JWST/NIRSpec data from \\ERO, GLASS and CEERS \\programmes}  \\ \midrule
\protect\cite{Curti2024MZR}           & $3-10$   &  [6, 10] & $0.17 \pm 0.03$ & $8.06\pm0.18$  & \makecell[l]{JWST/NIRSpec MSA data \\from the JADES programme. } \\
                                        &  $3-6$     & [6, 10]& $0.18\pm0.03$   & $8.11\pm0.03$ &  Uses calibrations from  \\
                                        &$6-10$    & [6, 10]& $0.11\pm0.05$   & $7.87\pm0.06$  &   \cite{Curti_FMR2020}\\ \midrule
\cite{Chemerynska2024}        & $\sim 7$ & [5.80, 7.19]& $0.39\pm0.02$   & $8.42 \pm0.17$ & \makecell[l]{JWST/NIRSpec low-\\resolution Prism spectra \\from UNCOVER Survey.} \\ \midrule
\cite{Chakraborty2025_MZR}        & $3-10$     & [7.5, 10] & $0.21\pm0.03$   & $7.99\pm0.21$ & \makecell[l]{ JWST/NIRSpec data}  \\ \midrule
\cite{2025arXiv250418616L}     & $\sim 6$ & [7.44, 8.84] & $0.26\pm0.01$   & $8.00\pm0.01$ &  \makecell[l]{$f_{[\ion{O}{iii}]\lambda5008}$ selected galaxies \\from NIRCam/WFSS  \\(ASPIRE and EIGER surveys)}  \\ \hline \midrule

SPHINX$^{20}$(No selection cut)   & $5-7$     & [6.51, 10.46]  &   0.41 &  8.10 &  Hydrodynamical simulation\\
 \hspace{1.45cm}(With selection cut)&$5-7$     & [8.46, 10.46] &  0.22  &  7.86 & \\  \midrule
FLARES\tablefootmark{b}  (No selection cut)    &  $5-7$      &   [8.0, 11.31]      &    0.44  & 8.25  &    Hydrodynamical simulation    \\
\hspace{1.42cm}(With selection cut)& $5-7$      &   [8.0, 11.31]     &   0.44 & 8.25   &  \\ \midrule
JAGUAR (No selection cut)    &  $5-7$      & [5.98, 9.96] &  0.24 &  8.41   & Catalogue from empirical relations \\ 
\hspace{1.45cm}(With selection cut) & $5-7$      & [6.86, 9.96] &  0.10  &   8.27 &\\ \bottomrule
\end{tabular}
}
\tablefoot{\\\tablefoottext{a}{Eq. \ref{eq: MZR} fit to data below turnover mass.}\\
\tablefoottext{b}{Eq. \ref{eq: MZR} fit to 50th percentile of the mass-weighted gas-phase metallicity.}}
\tablebib{
(1)SPHINX$^{20}$ \citep{2023OJAp....6E..44K}; (2)FLARES \citep{10.1093/mnras/staa3360, 10.1093/mnras/staa3715}; (3) JAGUAR catalogue \citep{2018ApJS..236...33W}; 
}
\end{table*}

\section{The relation between $[\ion{O}{iii}] \lambda 5008$ luminosity and stellar mass}
\label{Appendix: Simulations}
\begin{figure}[h]
\includegraphics[width=\columnwidth]{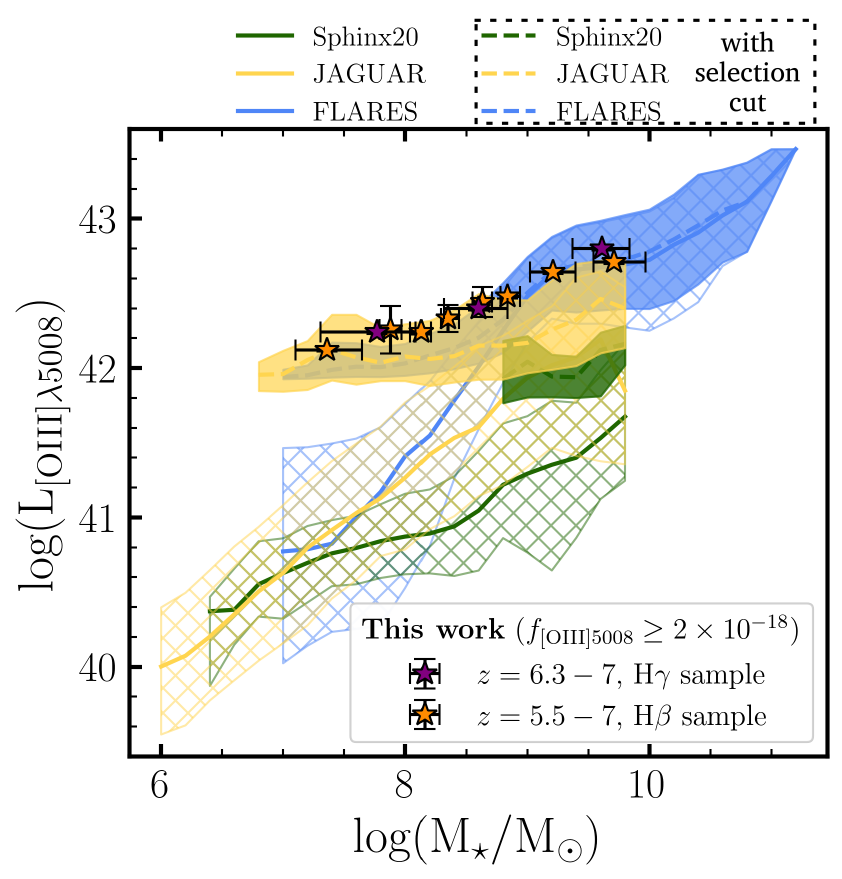}
\caption{$[\ion{O}{iii}]\lambda5008$ luminosities as a function of stellar mass for the SPHINX$^{20}$ \citep{2023OJAp....6E..44K} and FLARES simulation \citep{10.1093/mnras/staa3360, 10.1093/mnras/staa3715}, the JAGUAR catalogue \citep{2018ApJS..236...33W}. We plot both, the intrinsic relation and the relation after mimicking our selection cut. We also show the  $\rm L_{[\ion{O}{iii}]\lambda5008}- log(M_\star/M_\odot)$ relation for our stacked H$\gamma$ (purple stars) and H$\beta$ (orange stars) sample.}
\label{Sim_LO3}
\end{figure}

Before assessing the impact of our line-flux limited selection function on the mass - metallicity relation, it is important to compare the relation between the [\ion{O}{iii}] line luminosity and stellar mass among simulations and observations. In Fig. \ref{Sim_LO3}, we show [\ion{O}{iii}]$\lambda 5008$ luminosity as a function of stellar masses for H$\beta$ and H$\gamma$ sample, as well as for the hydrodynamical simulations SPHINX$^{20}$ and FLARES, and the JAGUAR catalogue. Upon applying our selection cut to these simulations, we find that the FLARES simulation and the JAGUAR catalogue match our relation between stellar mass and [\ion{O}{iii}] luminosities very well. The intrinsic relations are significantly steeper than observed, but this can be understood as being due to the selection cut. In SPHINX$^{20}$ however, only galaxies above stellar mass of $10^{8.4} \rm M_\odot$ are selected with the [\ion{O}{iii}] flux threshold. As mentioned in Sect. \ref{subsec: simulations}, SPHINX$^{20}$ has an elevated stellar mass - halo mass relation and therefore assigns too low SFRs at fixed mass, leading to much lower [OIII] luminosities at all masses.

\section{Derivation of the FMR equation used in our toy model} 
\label{Appendix: FMR_eq_model}
In in Sect. \ref{subsec: toy_models}, we describe our toy models used to understand the impact of the 3D relation between metallicity-stellar mass-SFR on the shape of the observed MZR. We described the 3D relation as,
\begin{equation}\label{eq: FMR eq_appendix}
    Z = Z_{\rm 0} + \rm M_{coeff} \times log_{10}(M_\star/M_\odot) + S_{coeff}\times  log_{10}(SFR/M_\odot yr^{-1})
\end{equation}

This can be expanded as,
\begin{flalign}
\begin{split}
    Z =  Z_{\rm 0} + \rm M_{ coeff} \times log_{10}(M_\star/M_\odot) + \\ S_{\rm coeff}\times \left( {\rm MS}_{\rm norm} + {\rm MS}_{\rm slope}\times \log_{10} \left(\frac{M_\star}{10^{10} M_\odot}\right) + \mathcal{N}(0,\sigma_{\rm SFR}) \right)
    \end{split}
\end{flalign}

Therefore,
\begin{flalign}
\label{eq: FMR eq exp}
\begin{split}
   Z =  Z_{\rm  0} + \rm (M_{\rm coeff} +S_{\rm coeff}\times {\rm MS}_{slope}) \times \log_{10}(M_\star/M_\odot) + \\ S_{\rm coeff}\times ({\rm MS}_{\rm norm} -10\times {\rm MS}_{\rm slope}) +  
   S_{\rm coeff}\times \mathcal{N}(0,\sigma_{\rm SFR})    
\end{split}
\end{flalign}
Here, we see that in the cases of $\rm S_{coeff} \neq 0$, the slope of the MZR regulated by parameters which  govern the influence of both stellar mass and SFR. The normalisation is dependent on the intrinsic metallicity floor and parameters governing the SFR. By design, the scatter is solely dependent in the parameters governing SFR. Therefore, in the presence of a 3D relation between stellar mass, metallicity and star formation rate, the shape of the MZR is governed by degenerate parameters. This makes it difficult to use Bayesian forward modelling techniques with our stacked measurements to obtain the intrinsic MZR.

\section{Effect of Selection function on MZR models}
\begin{figure*}
   \centering
   \begin{tabular}{ccc}
    \includegraphics[width=18cm]{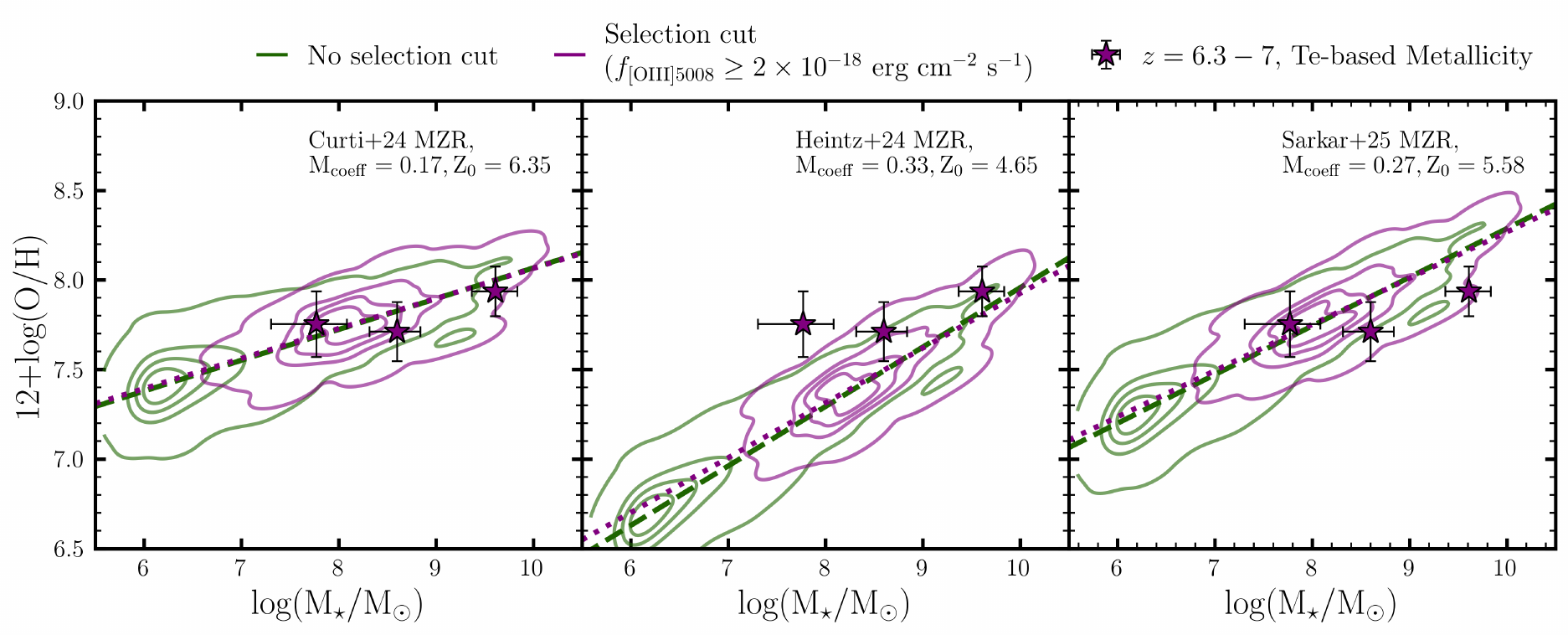}\\
     \includegraphics[width=18cm]{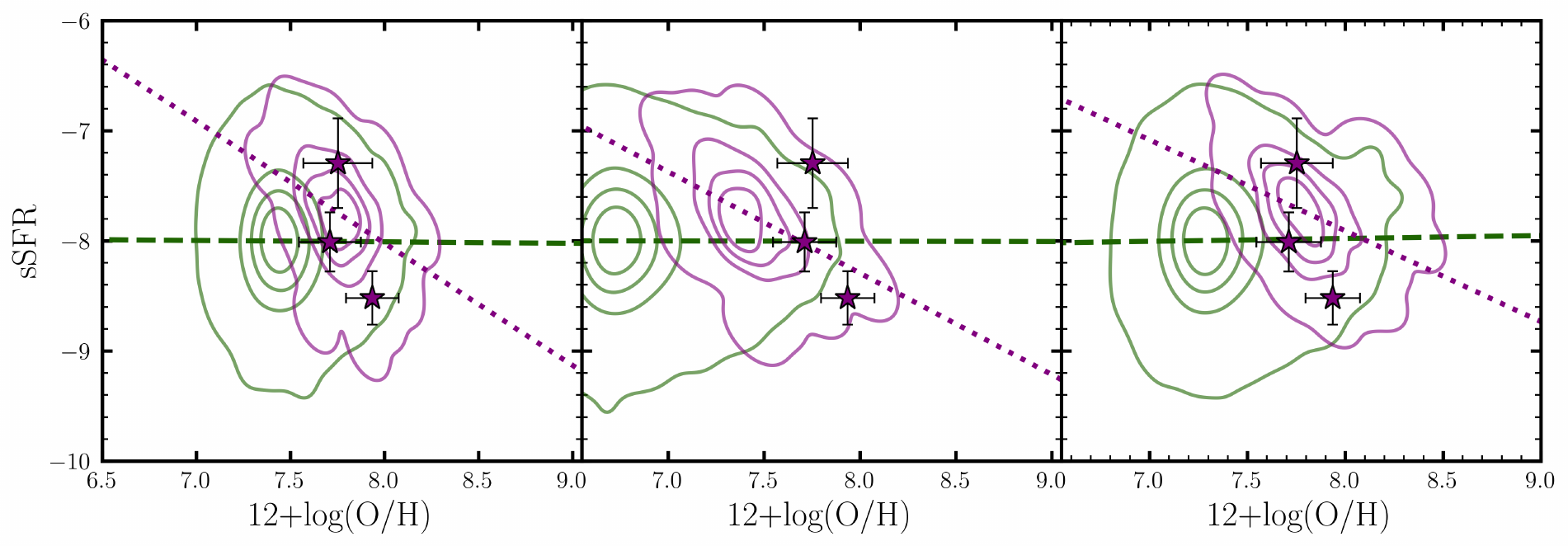}
    \end{tabular}
      \caption{(Top) The MZR for the `No SFR dependence' case toy model with three different MZR parameters. We also plot our direct-T$_e$ metallicity estimates. (Bottom) Specific star formation rate (sSFR) as a function of gas-phase metallicity. We show sSFR vs. gas-phase metallicity for our H$\gamma$ stellar mass binned stacks. The error bars on the sSFR are calculated using the 16th and 84th percentile in the stacks. The panels on the left are following the parameters by \cite{Curti2024MZR}, the middle panels follow the parameters by \cite{2023NatAs...7.1517H} and the panels on the right follow paramters by \cite{Sarkar2025_MZR} for Eq. \ref{eq: MZR}.}
         \label{fig: MZR_toy_models}
   \end{figure*}

We show the impact of our selection function on MZR parameters from \cite{2023NatAs...7.1517H} and \cite{Sarkar2025_MZR} in addition to the MZR parameters from \cite{Curti2024MZR} for case 2 ($\rm S_{coeff} = 0$) in Fig. \ref{fig: MZR_toy_models}. The scatter of the MZR is fixed to 0.1 dex for all three cases. Note that the scatter in the sSFR vs. 12+log(O/H) plane is different in the three cases even after fixing the scatter in the MZR plane. We find that our observations are in agreement with the `observed' contours on the sSFR vs. 12+log(O/H) plane despite the different parameters of the MZR. It is difficult to distinguish between the models because we only have three data points obtained our stacking analysis. To make a distinction, we require a large sample of individual measurements with an [OIII] flux cut selection to constrain the slope and the scatter.
\label{Appendix: MZR_models}
\end{appendix}

\end{document}